\documentclass[twocolumn,tighten,trackchanges]{aastex62}
\usepackage{graphicx}		
\usepackage{url}
\usepackage{epstopdf}
\usepackage{color}
\usepackage{multirow}
\usepackage{yhmath}
\usepackage{ragged2e}
\usepackage{hyperref}
\usepackage{url}
\usepackage{amsmath}
\DeclareGraphicsExtensions{.pdf,.png,.jpeg}

\DeclareMathOperator*{\argmin}{arg\,min}

\def\lsim{\mathrel{\raise.3ex\hbox{$<$\kern-.75em\lower1ex\hbox{$\sim$}}}}
\def\gsim{\mathrel{\raise.3ex\hbox{$>$\kern-.75em\lower1ex\hbox{$\sim$}}}}
\def\gtwid{\mathrel{\raise.3ex\hbox{$>$\kern-.75em\lower1ex\hbox{$\sim$}}}}
\def\proptwid{\mathrel{\raise.3ex\hbox{$\propto$\kern-.75em\lower1ex\hbox{$\sim$}}}}
\DeclareUnicodeCharacter{2212}{-}

\def\apjl{ApJL}
\def\apjs{ApJS}
\def\ao{Appl.~Opt.}
\def\aap{A\&A}
\newcommand{\ep}{\epsilon}
\newcommand{\suml}{\sum\limits}

\newcommand{\rel}{\mathcal{R}_{\rm rel}}
\newcommand{\abs}{\mathcal{R}_{\rm abs}}

\newcommand{\hati}{\widehat{I}}
\newcommand{\hatepsilon}{\widehat{\epsilon}}

\begin{document}

\title{THE ROLE OF ADAPTIVE RAY TRACING IN ANALYZING BLACK HOLE STRUCTURE}
\shorttitle{Adaptive Ray Tracing}
\author{Z. Gelles}
\affil{Center for Astrophysics $\vert$ Harvard \& Smithsonian, 60 Garden Street, Cambridge, MA 02138, USA}
\affil{Black Hole Initiative at Harvard University, 20 Garden Street, Cambridge, MA 02138, USA}
\author{B.~S.~Prather}
\affil{Department of Physics, University of Illinois at Urbana-Champaign, 1110 West Green Street, Urbana, IL 61801, USA}
\author{D.C.M. Palumbo}
\affil{Center for Astrophysics $\vert$ Harvard \& Smithsonian, 60 Garden Street, Cambridge, MA 02138, USA}
\affil{Black Hole Initiative at Harvard University, 20 Garden Street, Cambridge, MA 02138, USA}
\author{M.~D.~Johnson}
\affil{Center for Astrophysics $\vert$ Harvard \& Smithsonian, 60 Garden Street, Cambridge, MA 02138, USA}
\affil{Black Hole Initiative at Harvard University, 20 Garden Street, Cambridge, MA 02138, USA}
\author{G.~N.~Wong}
\affil{Department of Physics, University of Illinois at Urbana-Champaign, 1110 West Green Street, Urbana, IL 61801, USA}
\author{B. Georgiev}
\affil{Department of Physics and Astronomy, University of Waterloo, 200 University Avenue West, Waterloo, ON, N2L 3G1, Canada}
\affil{Waterloo Centre for Astrophysics, University of Waterloo, Waterloo, ON N2L 3G1 Canada}

\begin{abstract}
The recent advent of the Event Horizon Telescope (EHT) has made direct imaging of supermassive black holes a reality. Simulated images of black holes produced via general relativistic ray tracing and radiative transfer provide a key counterpart to these observational efforts. 
Black hole images have a wide range of physically interesting image structures, ranging from extremely fine scales in their lensed ``photon rings'' to the very large scales in their relativistic jets. The multi-scale nature of the black hole system is therefore suitable for a multi-scale approach to generating simulated images that capture all key elements of the system. 
Here, we present a prescription for adaptive ray tracing, which enables efficient computation of extremely high resolution images of black holes. Using the polarized ray-tracing code \texttt{ipole}, we image a combination of semi-analytic and GRMHD models, and we show that images can be reproduced with mean squared error of less than $0.1\%$ even after tracing $12{\times}$ fewer rays. We then use adaptive ray tracing to explore properties of the photon ring. We illustrate the behavior of individual subrings in GRMHD simulations, and we explore their signatures in interferometric visibilities.
\end{abstract}


\section{Introduction}
When surrounded by emitting material, black holes imprint distinctive properties of their spacetimes on the image seen by a distant observer. Black hole images can then offer valuable insights into the astrophysical processes that govern the accretion and outflow, the physical processes that produce heating and dissipation in the nearby plasma, and the geometrical lensing of light. Over the past few decades, images of black holes have evolved from being studied primarily for their rich theoretical features \citep{Luminet_1979,Bardeen_1973} to being directly accessible via very long baseline interferometry \citep{EHT_1,EHT_2,EHT_3,EHT_4,EHT_5,EHT_6}. With progressively sharper images of black holes expected as these observations continue to improve, increasingly accurate simulated images of black holes are imperative to guide analysis and interpretation. 

One limitation of image accuracy is related to finite image sampling at discrete points on the screen. Namely, the intensity at each point on an image is computed by ray tracing the path of the corresponding null geodesic and computing the radiative transfer along the trajectory. The computational expense of forming an image then increases with the number of rays at which this intensity function is sampled. A crucial question is how to efficiently distribute a finite sample of rays across an image to reach a prescribed image fidelity. 

Black hole ray-tracing programs typically distribute rays on an evenly spaced grid \citep[see, e.g.,][]{Gold_2020}. In this approach, regions of the image with sharp, bright features are sampled with the same density of rays as the faint regions of only diffuse structure. Black hole images are expected to have regions of both categories. Near the black hole, the accretion flow is turbulent and bright, requiring high resolution to adequately resolve. Far from the black hole, tightly collimated outflows or ``jets'' produce narrow regions with significant flux. The strong lensing of Kerr black holes is manifest in the ``photon ring," a bright ring with self-similar substructure that emerges in the limit of no absorption and scattering \citep[see, e.g.,][]{Luminet_1979,deVries_2000,Takahashi_2004,Beckwith_2005,Johannsen_Psaltis_2010,Gralla_2019,Johnson_2020}. Apart from these distinctive parts of the image, black hole images often have the bulk of their flux density concentrated in a small fraction of the image. 

In this paper, we develop a recursive scheme for black hole ray-tracing. We begin, in Section~\ref{sec:Background}, by summarizing previous work related to adaptive and high-resolution black hole imaging. Next, in Section~\ref{sec:Interpolation}, we describe expected interpolation errors in black hole images and present a recursive algorithm for efficiently generating high resolution images. In Section~\ref{sec:Results}, we evaluate the performance of our adaptive ray-tracing algorithm and explore the properties of extremely high-resolution features in the image and interferometric visibility domains. We consistently generate images with ${>}\,12{\times}$ fewer rays and ${<}~0.1\%$ mean squared error (MSE) compared to their truth counterparts. In Section~\ref{sec:apps}, we study the specific case of photon subrings in images from general relativistic magnetohydrodynamic (GRMHD) simulations and semi-analytic models, and we explore how different averaging prescriptions suppress stochastic image features. In Section~\ref{sec:Summary}, we summarize our results.

\section{Background and Literature Review}
\label{sec:Background}
Mathematical and computational techniques in general relativistic radiative transfer have seen tremendous developments during the past century \citep[see, e.g.,][]{Connors_1977,Rybicki_1979, Luminet_1979,Broderick_Blandford_2003,Broderick_Blandford_2004,Shcherbakov_2011,Gammie_2012,Krawczynski_2012,Beheshtipour_2017}. In order to use these techniques to render the appearance of black holes, the actual conditions of the surrounding plasma/emitting material are often simulated via GRMHD \citep[see, e.g.,][]{Gammie_2003, Narayan_2012}. Ray-tracing codes can then combine the output of GRMHD simulations with a radiative transfer scheme to produce black hole images under various astrophysical conditions \citep[see, e.g.,][]{Schnittman_2006,Noble_2007,Dolence_2009,Shcherbakov_2012}.

Among the GRMHD simulations, adaptive refinement techniques have been used extensively in the past for EHT-related work. Since GRMHD simulations evolve the accretion dynamics in individual ``cells" surrounding a black hole, codes can adaptively alter the size of these cells to better capture the relevant physical processes \citep{Porth_2019}. However, methods of spatial refinement in the subsequent ray-tracing analysis are not as well-documented, despite this method's ability to improve both image generation speed and quality.

Refinement strategies have already found some applications among ray-tracing programs. Many of the ray-tracing packages compared by \cite{Gold_2020} use adaptive step sizes to boost efficiency during numerical integration of the geodesic or radiative transfer equation \citep[see, e.g.,][]{Chan_2013,Dexter_2016,Pu_odyssey,Moscibrodzka2017ipole,Bronzwaer_2018}. 

\cite{Chan_2015} implemented a multi-scale sampling procedure by separately ray tracing images with uniform gridding but different pixel sizes and then combining them. In their three-layer scheme, each successive layer had the same number of pixels but increased the field of view by a factor of four, providing relatively high resolution near the black hole ($\Delta x = M/8$) and also giving a complete estimate for the X-ray flux from the simulated domain. 

Owing to their computational expense, high-resolution images of black holes are sparse in the growing literature related to black hole images. Notable exceptions include  \citet{Bronzwaer_2018}, who compared their \texttt{RAPTOR} code with \texttt{BHOSS} \citep{Younsi_2020} using images with $4096\times 4096$ pixels over a field of view of $60M$. Similarly, \citet{Davelaar_2018} used \texttt{RAPTOR} to generate a high-resolution, virtual reality simulation near a black hole with $2000\times 1000$ pixels per snapshot. 

To systematically produce high resolution images, especially with resolution to better resolve the substructure of the photon ring, a fully adaptive ray-tracing approach is advisable. Such a model has been implemented in the context of cosmological simulations \citep[e.g.,][]{Abel_2002,Wise_2011}. Our approach to adaptive ray-tracing more closely parallels the one implemented in the {\tt VRT}$^{\tt 2}$ package within {\sc THEMIS} \citep{Broderick_Blandford_2003,Broderick_Blandford_2004,Broderick_2020}, which chooses whether to ray-trace or interpolate on a pixel-by-pixel basis. Examples of high-resolution images with this adaptive ray tracing scheme can be seen in, e.g., \citet{Broderick_2006}. \citet{Parkin_2011} also presented an analogous refinement model for adaptive ray-tracing, although it was not developed for black hole images. \citet{Wong_Glimmer} presented an adaptive approach that uses only the path length of null geodesics, giving a spacetime-dependent grid that is independent of the image structure. We will next present a new method for adaptive ray tracing, which differs from these previous methods in its sampling methodology, and we will evaluate its performance using numerical simulations of black holes.

\section{Interpolation and Adaptive Image Refinement}
\label{sec:Interpolation}
The emission from a black hole produces a smooth intensity distribution on the screen of a distant observer, which we denote $I(\vec{x})$, where $\vec{x}$ is position. Ray tracing discretely samples $I(\vec{x})$ at a set of specified locations $\{ \vec{x}_i \}$. An output image $\hati(\vec{x})$ depends on the set of sampled rays $\{ \vec{x}_i \}$ and the interpolation method used to estimate a smooth distribution from this finite set. We now evaluate the expected interpolation errors for black hole images, and we present a recursive sampling approach that enables efficient estimates of high-resolution images.

\subsection{Continuous Error Approximations} 
\label{sec:continuouserror}
The simplest interpolation scheme to generate a full image from a discrete set of rays is a nearest neighbors approach. Given a rays sampled at $\vec{x}_1, \vec{x}_2,...,\vec{x}_n$, the nearest neighbors intensity distribution simply finds the closest sampled ray at every location: $\hati(\vec{x}) \equiv I\left( \argmin_{\vec{x}_i}{\left|\vec{x}_i-\vec{x} \right|} \right)$. 
A second interpolation scheme is a linear/bilinear approach, which is defined so that $\hati(\vec{x})$ is an average of its nearest surrounding rays that is weighted by distance. For a smooth intensity distribution, the errors in these two schemes are given by the first and second image spatial derivatives, respectively. 

Specifically, consider a point $\vec{x}$ lying equidistant from four rays spaced evenly around $\vec{x}$. Then, to leading order, the interpolation residuals will be (see Appendices \ref{sec:nearestapp} and \ref{sec:linearapp} for full derivation)
\begin{align}\label{eq:resideq}
     |\hati(\vec{x})-I(\vec{x})| &\sim \begin{cases}
     |\nabla \hati(\vec{x}) |\Delta x,&\text{Nearest}\\
     \frac{1}{4}|\nabla^2 \hati(\vec{x})|\Delta x^2 ,&\text{Linear}.
     \end{cases}
 \end{align}
where $\Delta x$ denotes the distance between $\vec{x}$ and each of its four surrounding rays.

\begin{figure*}[t]
    \centering
    \includegraphics[width=\textwidth]{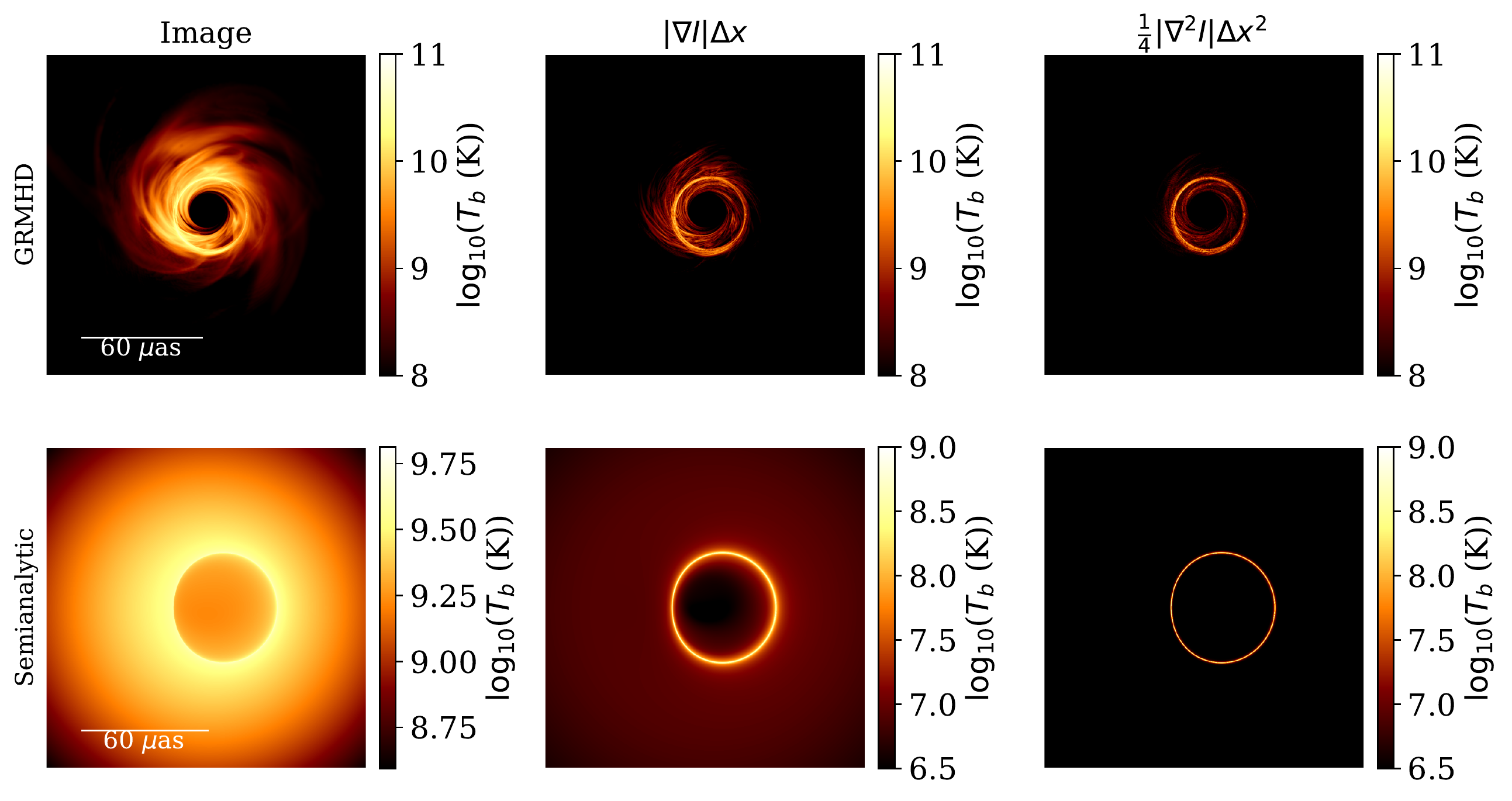}
    \caption{Simulated black hole images (left), their gradients (center), and their Laplacians (right). The top panels show a GRMHD snapshot, while the bottom panels show a semi-analytic model \citep[Test 1 of][]{Gold_2020}. All panels are rescaled to have units of brightness temperature (by multiplication by an appropriate power of $\Delta x\approx 0.156\, \mu{\rm as}$); the center and right panels then give the expected interpolation residual over this interval (Eq.~\ref{eq:resideq}).}
    \label{fig:gradpanelimage}
\end{figure*}

In Figure~\ref{fig:gradpanelimage}, we plot the gradients and Laplacians for ray-traced images of both a semianalytic model and a model from the EHT GRMHD library \citep{EHT_5}. In particular, the GRMHD model is a Magnetically Arrested Disk (MAD) \citep{Yuan_Narayan_2014} of M87*, with dimensionless spin $a_*=+0.94$, inclination angle $i=17^\circ$, and a field of view of $160\mu$as (corresponding to $\sim 44.17M/D$). We also use the electron heating parameter $R_{\rm high}=20$ \citep[for details, see][]{rhigh_2016,EHT_5}.

The semianalytic model mirrors Test 1 of \citet{Gold_2020} and models a spherically symmetric fluid distribution around a black hole located at a distance $D=7.78\,{\rm kpc}$ and with mass $M=4\times 10^6\,M_\odot$. Additional parameters for this semianalytic model include spin $a_*=0.9$, inclination angle $i=60^\circ$, and a field of view of ${\sim}152.289\,\mu{\rm as}$ ($30 M/D$).  For both models, we use a camera distance of $r_{\rm cam}=10^6M$ to mitigate small errors that may arise from the use of a pinhole camera. 

In all cases, the gradient and Laplacian of the intensity increase sharply near the photon ring because of strong gravitational lensing. Indeed, \citet{Psaltis_2015} proposed using sharp image gradients as a way to localize the photon ring; the Laplacian would also be an effective choice.

\subsection{Adaptive Ray-Tracing by Recursive Subdivision}
\label{sec:subdiv}
The highly localized regions of Figure~\ref{fig:gradpanelimage} with large derivatives suggest that judicious sampling can be used to significantly improve image generation speed and quality relative to a uniform grid. We will now describe an efficient recursive sampling procedure that is defined by a pair of error tolerances: $\abs$ and $\rel$.

To begin, we ray-trace an image on a grid at a coarse initial resolution, $n_{0x}\times n_{0y}$, where $n_{0x}$ and $n_{0y}$ are any two integers. Next, we selectively populate an image with finer resolution by a factor of two: $(2n_{0x}-1)\times (2n_{0y}-1)$. For each point, we either ray trace or interpolate based on the expected relative and absolute interpolation errors:
\begin{align}\label{eq:errordefrel}
    \ep_{\rm rel}(\vec{x})&=\left|\frac{I(\vec{x})-\hati(\vec{x})}{I(\vec{x})}\right|,\\
        \label{eq:errordefabs}
    \ep_{\rm abs}(\vec{x})&=\left|\frac{I(\vec{x})-\hati(\vec{x})}{\overline{I}}\right|.
\end{align}
Both $\ep_{\rm abs}$ and $\ep_{\rm rel}$ are dimensionless, but they differ in their normalization: $\ep_{\rm rel}$ uses the local image intensity $I(\vec{x})$, while $\ep_{\rm abs}$ uses the image-averaged intensity $\overline{I}$. If $\ep_{\rm abs} > \abs$ and $\ep_{\rm rel} > \rel$, then $I(\vec{x})$ is computed by ray-tracing. Otherwise, $I(\vec{x})$ is computed by interpolation. Hence, rays are computed only where interpolation residuals are expected to be large. This procedure can then be repeated arbitrarily, giving an effective final image resolution that is $n_{x}\times n_{y}$, with 
\begin{align}
  n_x&=2^N(n_{0x}-1)+1\\
\nonumber  n_y&=2^N(n_{0y}-1)+1.
\end{align}

\begin{figure}[th]
    \centering
    \includegraphics[width=\columnwidth]{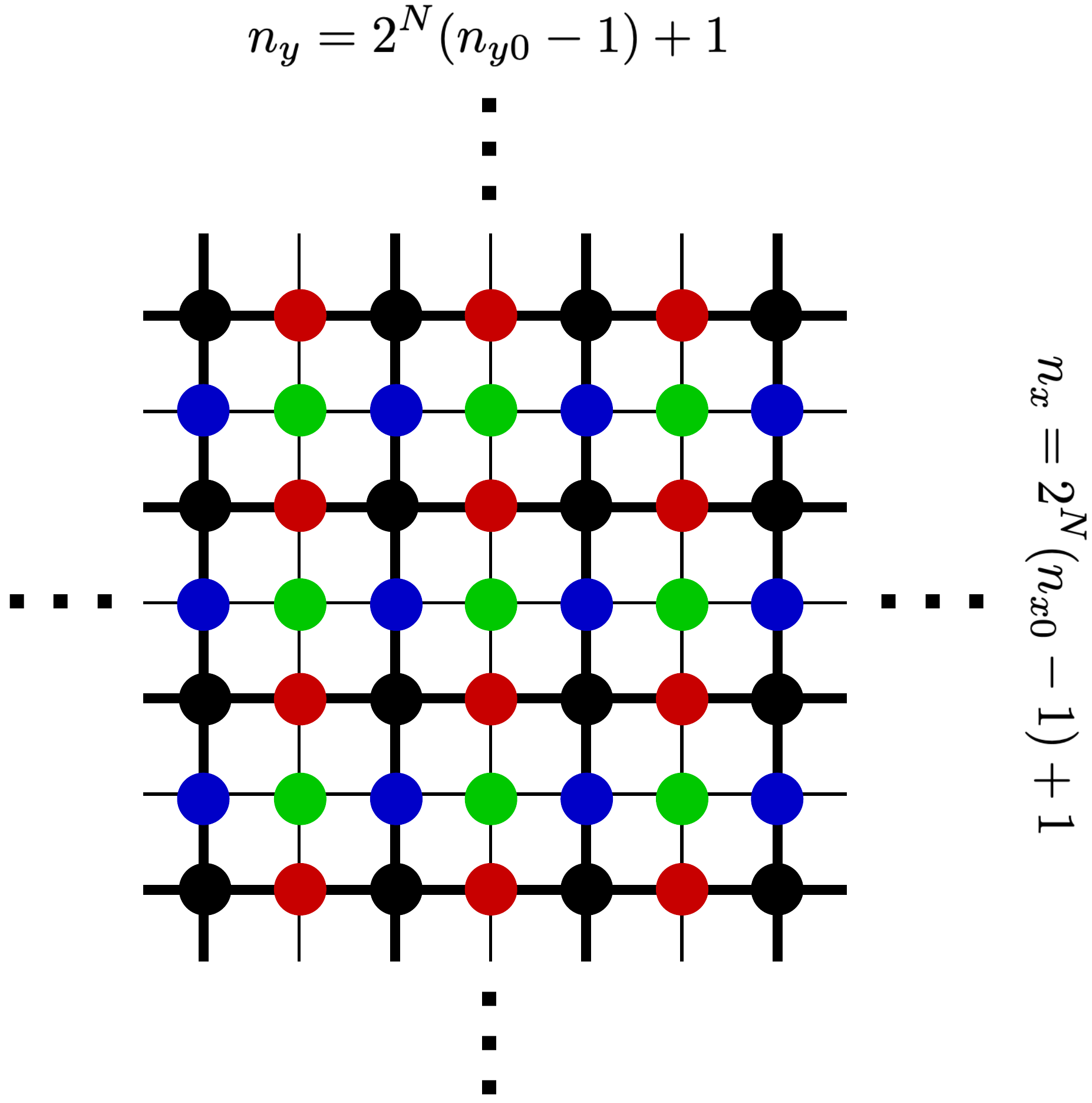}
    \caption{Recursive gridding approach. Black dots show points sampled at the $n^{\rm th}$ refinement level; colored dots show additional points sampled at the $(n+1)^{\rm th}$ refinement level. Distinct colors represent distinct cases for interpolation error estimation.}
    \label{fig:pixeldrawing}
\end{figure}

\begin{figure}[t]
    \centering
    \includegraphics[width=0.8\columnwidth]{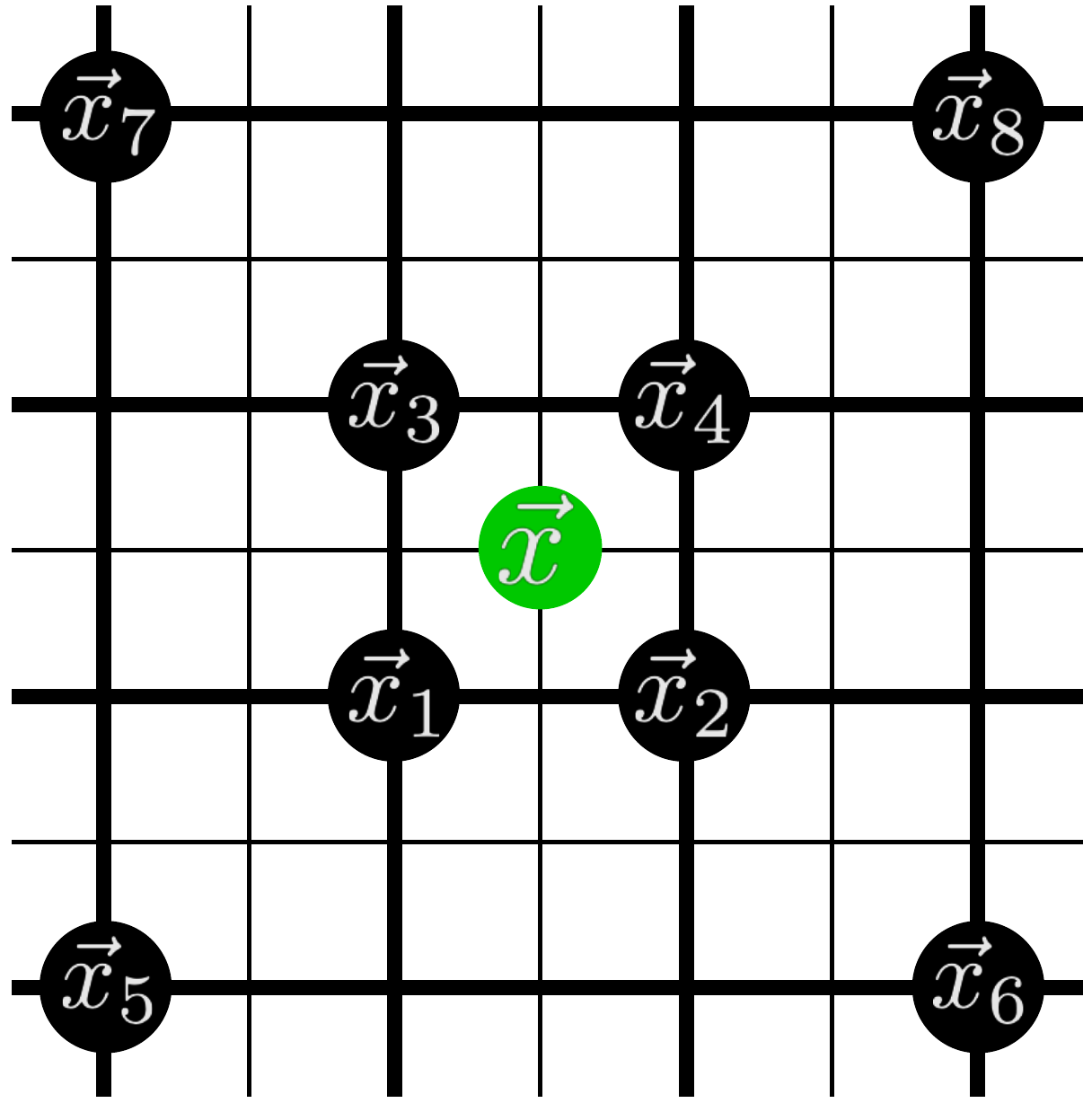}
    \caption{Corner pixels used to compute the interpolation uncertainties $\ep_{\rm abs}(\vec{x}$) and $\ep_{\rm rel}(\vec{x})$. For nearest neighbor interpolation, only points $1-4$ are needed, while linear interpolation requires points $5-8$ as well.}
    \label{fig:pixeldrawingzoom}
\end{figure}

Our recursive approach relies on computing estimates $\hatepsilon_{\rm rel}(\vec{x})$ and $\hatepsilon_{\rm rel}(\vec{x})$ for the interpolation residuals, and these estimates change for different configurations of pixels. As shown in Figure \ref{fig:pixeldrawing}, each point $\vec{x}$ lying on the $(n+1)^{\rm th}$ grid will fall into one of four categories:\begin{itemize}
    \item Category 1 (Black): $\vec{x}$ had its intensity computed at the $n^{\rm th}$ refinement level; there is no interpolation residual.
    \item Category 2 (Blue): $\vec{x}$ lies in between two points located above and below, each of whose intensities were computed at the $n^{\rm th}$ refinement level.
    \item Category 3 (Red): $\vec{x}$ lies in between two points located to left and right, each of whose intensities were computed at the $n^{\rm th}$ refinement level.
    \item Category 4 (Green): $\vec{x}$ lies in between four equidistant corner points, each of whose intensities were computed at the $n^{\rm th}$ refinement level.
\end{itemize} 
For cases 2--4, we use finite differences to estimate derivatives and then take the appropriate leading term in the Taylor series approximation to evaluate interpolation uncertainties (see, e.g., Appendix A of \citealp{finitedifferences}). For example, consider estimating the intensity at the central point of Figure~\ref{fig:pixeldrawing}, whose location we will call $\vec{x}$. Defining $\hati_{i}\equiv \hati(\vec{x}_i)$ and labeling the points as in Figure~\ref{fig:pixeldrawingzoom}, we find:
\begin{align}
\label{eq:epabs}
    \hatepsilon_{\rm abs}(\vec{x}) &= \begin{cases}
     \left|\frac{1}{\overline{I}_{\rm int}}\sqrt{\left(\frac{\hati_4-\hati_1}{2}\right)^2+\left(\frac{\hati_3-\hati_2}{2}\right)^2}\right|,&\text{Nearest}\\~\\
    \left| \frac{-(\hati_1+\hati_2+\hati_3+\hati_4)+(\hati_5+\hati_6+\hati_7+\hati_8)}{16\overline{I}_{\rm int}}\right| ,&\text{Linear},
     \end{cases}
     \\
     \label{eq:eprel}
    \hatepsilon_{\rm rel}(\vec{x}) &= \begin{cases}
     \left|\frac{1}{\hati_1}\sqrt{\left(\frac{\hati_4-\hati_1}{2}\right)^2+\left(\frac{\hati_3-\hati_2}{2}\right)^2}\right|,&\text{Nearest}\\~\\
    \left| \frac{-(\hati_1+\hati_2+\hati_3+\hati_4)+(\hati_5+\hati_6+\hati_7+\hati_8)}{4(\hati_1+\hati_2+\hati_3+\hati_4)}\right| ,&\text{Linear}.
     \end{cases}
\end{align}
Here, the nearest neighbor is chosen (arbitrarily) to be $\hati_1$. Additionally, $\overline{I}_{\rm int}$ is defined as the interpolated average intensity after the first pass of ray-tracing (with resolution $n_{0x}\times n_{0y}$). For a detailed derivation of Equations~\ref{eq:epabs} and \ref{eq:eprel}, see Appendix~\ref{sec:nearestapp} and \ref{sec:linearapp}.

While we have derived estimates for $\hatepsilon_{\rm abs}(\vec{x})$ and $\hatepsilon_{\rm rel}(\vec{x})$ using the simplified approximation of a smooth image with interpolation residuals dominated by low-order derivatives, these estimates are also useful for the more general case of images with small-scale structure and sharp gradients. For example, if the small-scale image structure is stochastic with a power-law spectrum, then we can compare the ensemble-average properties of the exact (\autoref{eq:errordefrel}) and estimated (\autoref{eq:epabs}) interpolation residuals. For 1D interpolation, we find
\begin{align}
    \frac{\left \langle \hatepsilon_{\rm abs}^2 \right \rangle^{\frac{1}{2}}}{\left \langle \ep_{\rm abs}^2 \right \rangle^{\frac{1}{2}}} 
    &= 
    \begin{cases}
    2^{\beta/2-2},&\text{Nearest}\\~\\
    2^{\frac{\beta}{2}-4} \sqrt{\frac{1 + 2^{\beta-1} - 3^{\beta-2}}{1 - 2^{\beta-4}}} ,&\text{Linear},
     \end{cases}
\end{align}
where $\beta$ is the power-law exponent. GRMHD simulation often have $\beta \sim 2.5$, giving ratios ${\left \langle \hatepsilon_{\rm abs}^2 \right \rangle^{\frac{1}{2}}}/{\left \langle \ep_{\rm abs}^2 \right \rangle^{\frac{1}{2}}} $ of 0.59 and 0.27 for 1D nearest and linear interpolation, respectively. Thus, even in this generalized case of stochastic image fluctuations with a power-law spectrum (for which a local Taylor series expansion is poor), our approximate estimates for interpolation error will still provide useful refinement criteria. For additional discussion of the interpolation errors for images with power-law spectra, see \autoref{sec:powerlaw_spectrum}.

Previous uses of adaptive ray-tracing have used more complicated refinement criteria than the one we have selected. \citet{Parkin_2011} also implemented a second order criterion (the \citealp{Lohner_1987} criterion), which uses multiple cross derivatives. Similarly, {\tt VRT}$^{\tt 2}$ employs a bicubic interpolator. However, we expect the benefit of these higher-order schemes to be marginal in the case of an image with a turbulent power spectrum (see Appendix~\ref{sec:powerlaw_spectrum}). For the remainder of this paper, we use the linear interpolation scheme. 

\section{Results}
\label{sec:Results}

\subsection{Models and Implementation}
We implement the recursive ray-tracing scheme described above into the polarized general relativistic radiative transfer (GRRT) code \texttt{ipole}\footnote{https://github.com/moscibrodzka/ipole. Our adaptive tools were developed at https://github.com/AFD-Illinois/ipole.} \citep{Moscibrodzka2017ipole}. This implementation allows us to assess the performance of our algorithm and subsequently generate extremely high-resolution images of black holes. 

In addition to the MAD model and spherical semianalytic model used in Figure~\ref{fig:gradpanelimage}, we also generate images using a Standard and Normal Evolution (SANE) model from the EHT GRMHD library \citep[][]{Yuan_Narayan_2014,EHT_5}, as well as a semianalytic model of a geometrically thin, rotating disk (Test 5 of \citealp{Gold_2020}).

Before generating images, however, we must first determine suitable error tolerances $\abs$ and $\rel$. To evaluate the effect of different tolerances on the resultant image, we use the following three metrics (the second and third of which are also used in  \citealp{Gold_2020}):\begin{align}
    \text{Interpolation Fraction}&\equiv \frac{\text{\# pixels interpolated}}{n_x\times n_y},
    \label{eq:interpdef}
    \\
    \text{Flux Error}&\equiv\left|\frac{F-\widehat{F}}{F}\right|,
    \label{eq:fluxdef}
    \\
    \text{Mean Squared Error}&\equiv \frac{\suml_{{\rm pixels}\, i}|I(\vec{x}_i)-\hati(\vec{x}_i)|^2}{\suml_{i}|I(\vec{x})|^2}.
    \label{eq:msedef}.
\end{align}
Here, $F\equiv \int {\rm d}^2 \vec{x} \,I(\vec{x})$ is the image flux associated with the model intensity distribution, and $\widehat{F}\equiv \int {\rm d}^2 \vec{x} \,\hati(\vec{x})$ is the image flux associated with the estimated intensity distribution. We note that while a high flux error necessarily implies a high MSE, we include the former metric since it has a clear physical interpretation and bears more relevance in potential applications to light curves.

\begin{figure*}[t]
    \centering
    \includegraphics[width=.48\textwidth]{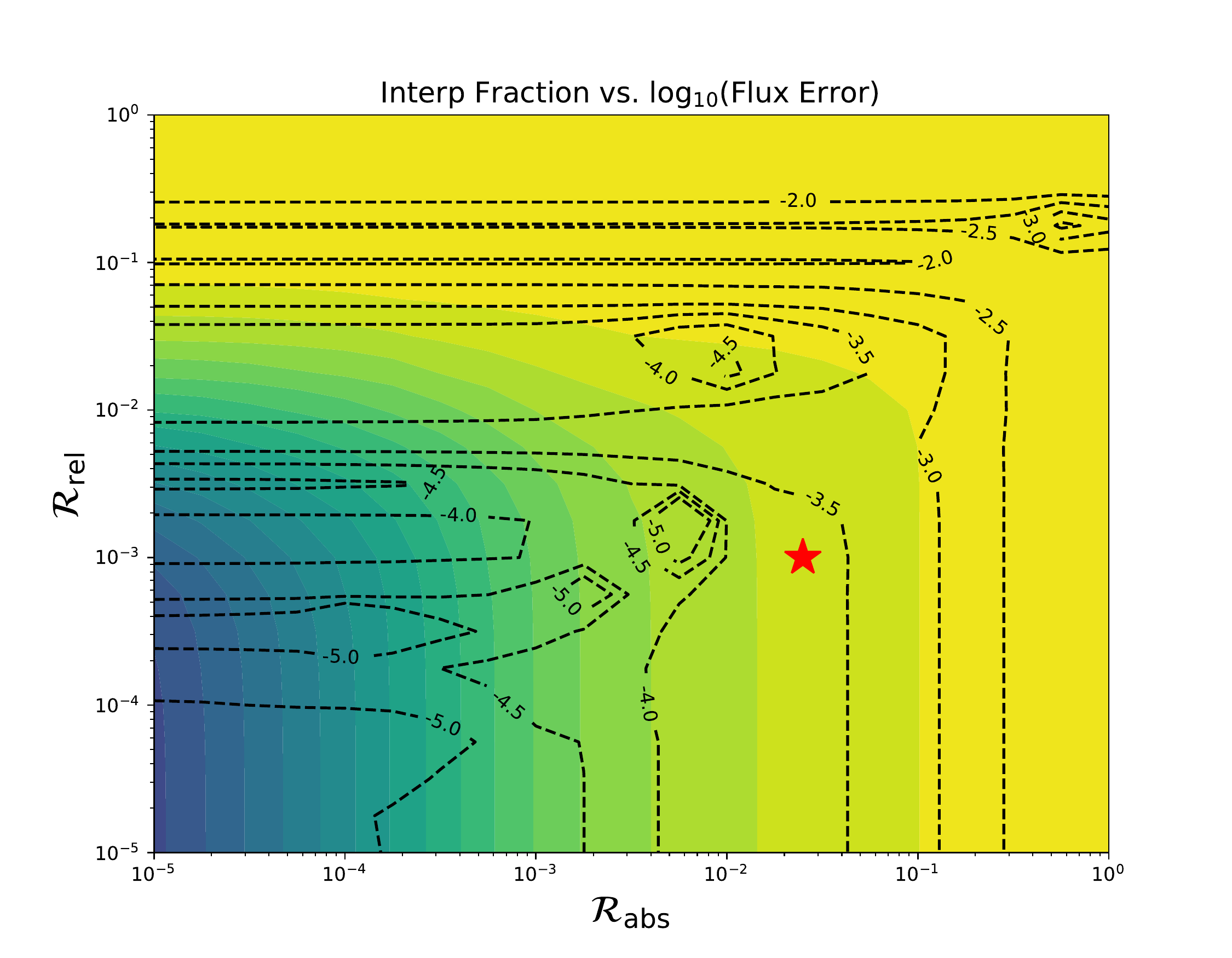}
    \includegraphics[width=.51\textwidth]{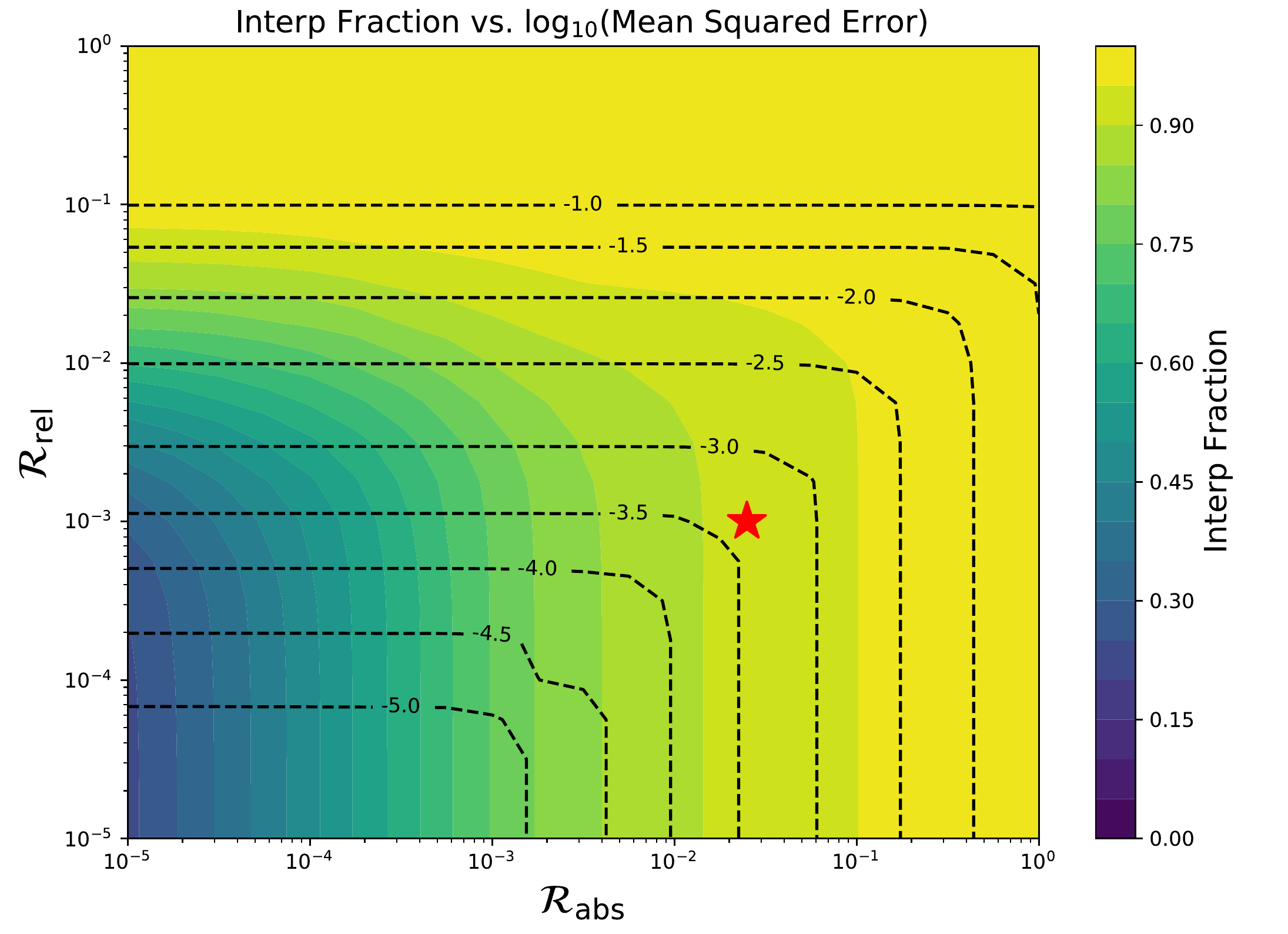}
    \caption{Contour plots representing how the error metrics depend on the tolerances $\rel$ and $\abs$ for the MAD GRMHD snapshot traced with $1025 \times 1025$ pixels on $160\,\mu{\rm as}$ FOV. Colors represent interpolation fraction while dashed lines represent the error (flux error on left and mean squared error on right). The star indicates the error tolerances we choose to generate the subsequent GRMHD images in this paper $(\rel=0.001$ and $\abs = 0.025$).}
    \label{fig:contours}
\end{figure*}

Our goal is to maximize the interpolation fraction (IF) while minimizing the flux error (FE) and MSE, as this achieves both efficiency and accuracy. To this end, we run our adaptive scheme on a single GRMHD snapshot (the same snapshot/parameters used in Figure \ref{fig:gradpanelimage}) with a wide variety of error tolerances, and we evaluate the IF/FE/MSE for each. The contours in Figure \ref{fig:contours} show how the error metrics each depend on $\rel$ and $\abs$. For this comparison, we take $n_{0x}=n_{0y}=65$ pixels and $n_x=n_y=1025$ pixels, and we compute the model intensity distribution $I(\vec{x})$ by ray-tracing  each pixel on a $1025 \times 1025$ grid. 

As expected, as $\rel$ and $\abs$ increase, the adaptively sampled image $\hati(\vec{x})$ increasingly deviates from the fully sampled ${I}(\vec{x})$. 
As  $\rel \rightarrow \infty$ and $\abs \rightarrow \infty$, all adaptive refinement will cease and the image will be given by the initial sampling. For the initial $65 \times 65$ sampling of this snapshot,  ${\rm MSE} \approx 0.22\approx 10^{-0.65}$ and ${\rm FE} \approx 0.029\approx 10^{-1.54}$. These serve as reference values to compare against iterations of the scheme with more stringent tolerances.

\begin{figure*}[t]
    \centering
    \includegraphics[width=\textwidth]{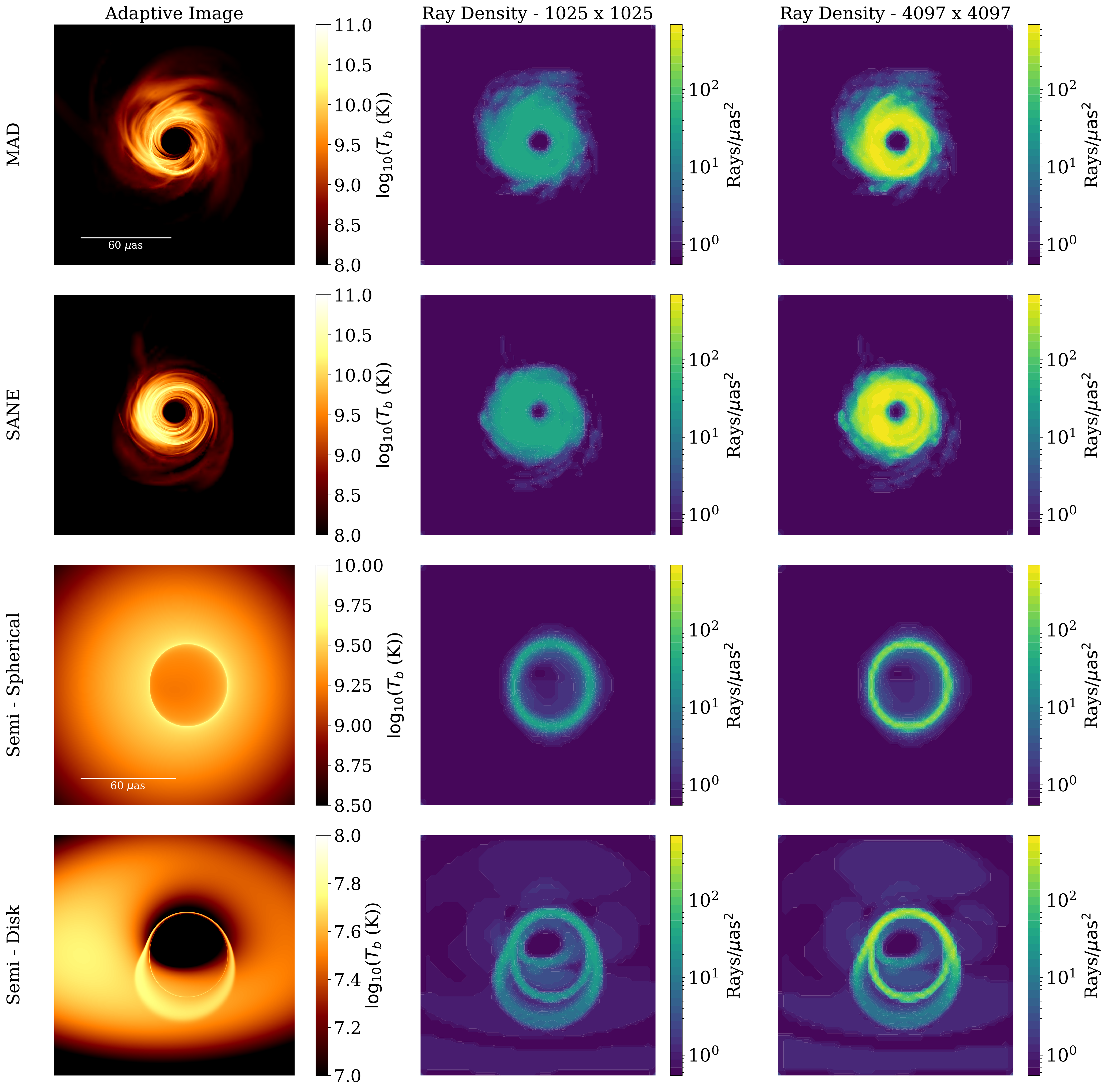}
    \caption{Adaptively ray-traced images, shown with ray-density plots to the right. Rays are concentrated near the photon ring, with occasional rays scattered across the diffuse emission. To compute the ray density, we use a kernel size equal to the spacing between pixels at the zeroth refinement level: $17\times 17$ pixels for the lower resolution image and $65\times 65$ pixels for the higher resolution image.}
    \label{fig:adaptiveimages}
\end{figure*}

The contours in Figure \ref{fig:contours} can be used for practical image generation purposes -- for a desired accuracy, one can select $\abs$ and $\rel$ such that the interpolation fraction and hence the efficiency are maximized. For our subsequent analysis of GRMHD images, we choose $\abs=0.025$ and $\rel=0.001$ (the red star in Figure \ref{fig:contours}), as this set of tolerances allows for ${>}\,90\%$ interpolation while constraining the MSE and FE to ${<}\,0.1\%$.

For subsequent analysis of the semianalytic models, we choose $\abs=\rel=0.001$, as we find that the semianalytic models are slightly more sensitive to the absolute tolerance. 

\subsection{High-Resolution Images and Analysis}
\label{sec:highres}
Using the tolerances selected described above, we adaptively ray-trace in $\texttt{ipole}$ at resolutions of $1025\times 1025$ and $4097\times 4097$ pixels. Images of the four models for the latter resolution are shown in Figure \ref{fig:adaptiveimages}, along with their respective ``ray density plots," illustrating where the program is concentrating rays in each image. The ray density plots are generally consistent with Figure \ref{fig:gradpanelimage} -- the majority of ray-tracing takes place in an annulus about the photon ring, while the background and shadow are predominantly interpolated. As with Figure \ref{fig:gradpanelimage}, the annuli for the semianalytic models are significantly thinner and more pronounced.

\begin{figure*}[t]
    \centering
    \includegraphics[width=\textwidth]{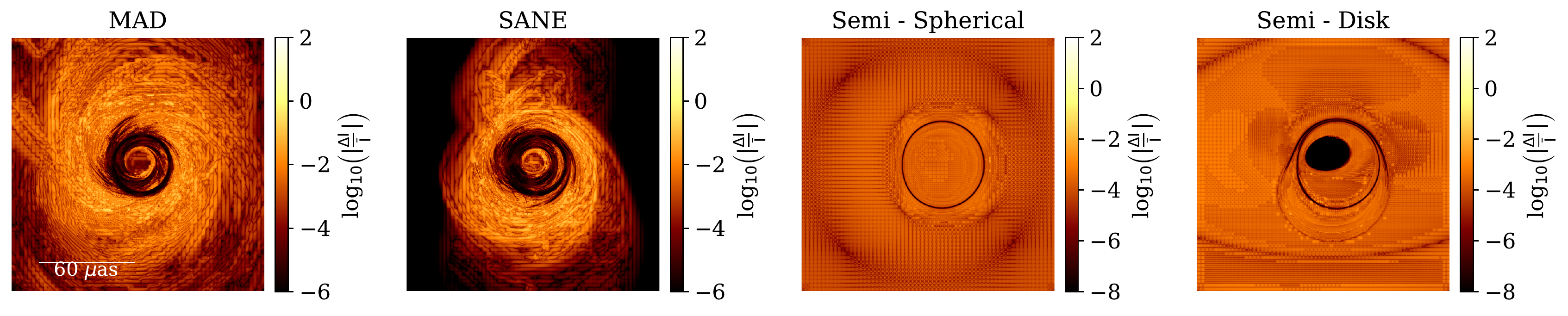}
    \caption{Normalized residual images for the $4097\times4097$ snapshots of the GRMHD models and semianalytic models. Regions with a high concentration of rays have particularly small residuals, as do regions with diffuse structure. For reference, the image-averaged intensities $\overline{I}$ for the four images are (after converting to brightness temperature): $5.2\times 10^8$\,K, $7.0\times 10^8$\,K, $1.6\times 10^9$\,K, and $2.5\times 10^7$\,K, from left to right.}
    \label{fig:residimage}
\end{figure*}

The corresponding residual images (fully ray-traced images subtracted from adaptively ray-traced images and normalized by the image-averaged intensity: $\left| \frac{\hati(\vec{x}) - I(\vec{x})}{\overline{I}}\right|$) are shown in Figure~\ref{fig:residimage}. In all images, the residual amplitudes are relatively constant, which reflects refinement goal of $\ep_{\rm abs}$. In portions of the image that correspond to rays that have been traced, the residuals are zero. In particular, all images contain a base grid of $65\times 65$ evenly spaced rays, giving evenly spaced nulls in the residuals that align with this grid. This feature is most evident in the semianalytic models, where the smooth intensity distribution allows significant interpolation.

\begin{table}
\footnotesize
 \begin{tabular}{ |p{2.08cm}||p{.48cm}|c|c|c|}
 \hline
 Model - $1025^2$ & \,\,IF  & FE & MSE& \# Rays \\
 \hline
 {\scriptsize GRMHD - MAD}  &  0.92 & $2.5\times 10^{-4}$ & $4.6\times 10^{-4}$ & $8.7\times 10^4$\\
 {\scriptsize GRMHD - SANE}  & 0.92   & $2.3\times 10^{-4}$ & $2.5\times 10^{-4}$& $9.0\times10^4$\\
{\scriptsize Semi - Spherical} & 0.97  & $1.5\times 10^{-5}$ & $5.6\times 10^{-7}$&  $2.9\times 10^4$\\
{\scriptsize Semi - Disk}    &  0.96  & $1.5\times 10^{-4}$ & $1.3\times 10^{-5}$&$4.3\times10^4$ \\

 \hline
\end{tabular}

\begin{tabular}{ |p{2.08cm}||p{.48cm}|c|c|c|}
 \hline
 Model - $4097^2$ & \,\,IF & FE & MSE & \# Rays\\
 \hline
{\scriptsize GRMHD - MAD}   & 0.96  & $4.2\times 10^{-4}$ & $1.5\times10^{-4}$ & $7.1\times 10^5$\\
{\scriptsize GRMHD - SANE} & 0.95 & $2.5\times 10^{-4}$ & $8.3\times 10^{-5}$ & $7.9\times10^5$\\
{\scriptsize Semi - Spherical} &  0.99 & $2.1\times 10^{-5}$ & $1.8\times 10^{-7}$&$1.0\times 10^5$ \\
 {\scriptsize Semi - Disk}    & 0.99 & $1.3\times 10^{-4}$ & $3.7\times 10^{-6}$ & $1.9\times 10^5$ \\
 \hline
\end{tabular}
\caption{Statistics for $1025\times 1025$ and $4097\times 4097$ images. Here, the FE and MSE give the values for an adaptively ray-traced image at the specified resolution relative to a fully ray traced image at that resolution. The last column shows the number of rays traced in the simulation. All GRMHD images used $\rel=0.001$ and $\abs=0.025$, while semianalytic images used $\rel=\abs=0.001$.} 
\label{tab:imagestats}
\end{table}

The resultant error statistics (Equations \ref{eq:interpdef}--\ref{eq:msedef}) for these adaptively ray-traced images are presented in Table \ref{tab:imagestats}. The $1025\times 1025$ images match the predictions from the contours in Figure \ref{fig:contours}: both the MSE and FE are lower than $0.1\%$, while the interpolation fraction exceeds $90\%$. In Table~\ref{tab:imagestats}, we also list the total number of rays traced in each image: $(1-{\rm IF})\times n_x\times n_y$, which is roughly proportional to computational expense.
 
While these statistics were generated for a specific choice of image parameters, we obtain similar results for other models. We ray-trace with the same tolerances on all combinations of MAD/SANE, $a_*=-0.94, 0,+0.94$, and $R_{\rm high}=20,\, 40,\, 80$. We find that among these images, IF ranges from 0.877 to 0.929, FE ranges from $3.02\times 10^{-5}$ to $8.59\times 10^{-4}$ and MSE ranges from $2.82\times 10^{-4}$ to $5.55\times 10^{-4}$.

We may further vary the magnetization $(\sigma)$ cut \citep[see, e.g.,][]{Chael_2019}, which is a quantity designed to restrict emission to regions where the fluid density has not been physically invalidated. Regenerating the MAD image from Figure \ref{fig:gradpanelimage} with $\sigma=1,5,10,\,{\rm and}\, 50$, we find that IF ranges from $0.917$ to $0.931$, FE ranges from $2.50\times 10^{-4}$ to $7.50\times 10^{-4}$, and MSE ranges from $2.79\times 10^{-4}$ to $4.69\times 10^{-4}$. Thus, for these GRMHD simulations, the accuracy and efficacy of the adaptive refinement is relatively insensitive to the choice of accretion and radiation model parameters. 

The semianalytic models have  lower errors than the GRMHD models and higher interpolation fractions. This is likely because the smooth underlying structure in semianalytic models is better suited to interpolation. 

\begin{figure}[t]
    \centering
    \includegraphics[width=0.49\textwidth]{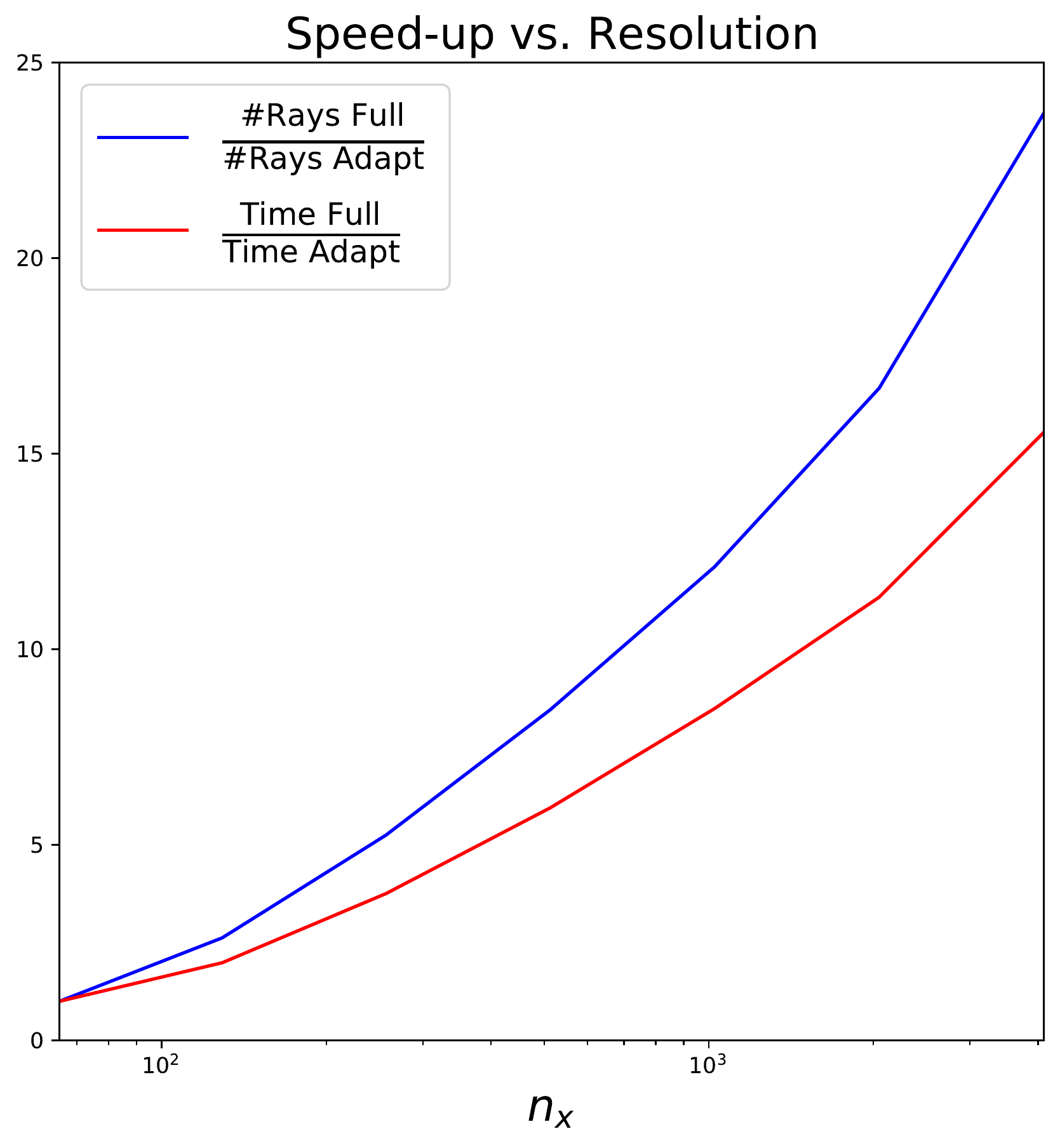}   
    \caption{Adaptive ray-tracing speed-up factor as a function of image resolution for the MAD GRMHD model shown in Figure~\ref{fig:gradpanelimage}. The red line shows the full-to-adaptive ratio of runtime, while the blue line shows the full-to-adaptive ratio of the number of rays. This latter quantity represents the idealized speedup factor if all pixels took equally long to ray-trace. Here, ``full" refers to an image that was ray-traced at every pixel (i.e., $\hati(\vec{x}) = I(\vec{x})$ at all pixels on the image).}
    \label{fig:speedup}
\end{figure}

While IF is a useful metric to quantify sampling efficiency, it does not directly correspond to the reduction in image generation time. 
Namely, rays near the photon ring are more expensive to trace, as their geodesics have longer paths through the emitting material and thus require more calculations to perform the radiative transfer. Nevertheless, even though our adaptive scheme predominantly samples in this computationally expensive region, we find that the reduction in image generation time is similar to the interpolation fraction (see Figure~\ref{fig:speedup}). In general, the relationship between interpolation fraction and computational expense will depend on details of the underlying model and of the GRRT implementation.

We note additionally from Table \ref{tab:imagestats} that as resolution increases, IF increases as well. 
This behavior is expected because the error estimates decrease at smaller separations, allowing more pixels to satisfy the error tolerances required for interpolation. 
Indeed, Figure~\ref{fig:adaptiveimages} shows that the density of rays in the diffuse parts of the image quickly saturates -- upon increasing the spatial resolution from $1025\times 1025$ to $4097\times 4097$, the sampled ray density only increases in the photon ring region. 

Unlike the IF, however, the MSE and FE do not depend strongly on resolution. 
These quantities instead depend on the tolerances $\rel$ and $\abs$. Thus, even though more pixels are interpolated at a resolution of $4097\times 4097$ compared to a resolution of $1025\times 1025$, the output images have comparable accuracy by these two metrics. 

\subsection{Visibility Domain Analysis}
VLBI directly measures interferometric visibilities, which correspond to complex Fourier components of the sky image \citep{TMS}. 
Hence, visibility-domain tests are appropriate to assess suitability for direct comparisons with observables. 
In this section, we analyze the visibility spectra of our adaptively ray-traced images and show that they exhibit the universal properties expected for black hole images.

\subsubsection{Expected Visibility Signatures}
\label{sec:vissig}

For all images (both GRMHD and semianalytic), we expect the visibility spectra to reflect signatures of the strong gravitational lensing of light. Namely, in the Kerr spacetime, photons can complete spherical orbits at a fixed set of Boyer-Lindquist radii, and null geodesics near these orbits approach a ``critical curve" on an observer's screen upon eventual escape \citep{Bardeen_1973,Teo_2003}. Rays that terminate increasingly near the critical curve make increasingly many revolutions around the black hole, producing a bright ``photon ring'' when the emitting material is optically thin.  
The visibility spectrum of such a ring should exhibit damped oscillatory behavior with a characteristic period of $1/d$, where $d$ is the photon ring's screen diameter projected along the baseline direction \citep{Johnson_2020}. 

A photon that ends up within the photon ring may be further labelled by a number $n$ representing the number of half-orbits the photon has taken around the black hole between emission and reception at the observer.
In the case of a geometrically thin disk of emitting plasma, the photon ring naturally decomposes into a series of overlapping, self-similar ``subrings''  indexed by $n$, where each subring comprises the set of all photons labelled $n$. Because the subrings are exponentially demagnified, the flux from each successive subring will dominate the visibility spectrum for a range of baselines that sample angular scales matched to those of the subring \citep{Johnson_2020}. 

In addition to the effects of gravitational lensing, we also expect to identify the signature of accretion turbulence present in GRMHD images. These will produce stochastic visibility noise that may be described by their power spectrum (i.e., the squared visibility amplitude). 


\subsubsection{Visibility Spectra of High-Resolution Images}

We now explore the expected long-baseline visibility signatures using our high-resolution images computed with adaptive ray tracing. 
While short interferometric baselines will have a complex visibility structure that depends on the overall image morphology, the visibility signatures from the photon ring and from turbulence will emerge on baselines that heavily resolve the image. Specifically, to accurately estimate the visibility on a baseline with dimensionless length $u$ requires angular resolution $\Delta\theta\sim\frac{1}{u}$ \citep{TMS}. 
For our images of both the GRMHD and semianalytic models, the visibility spectra can thus be computed to baselines of $u_{\rm max}\gtrsim 5000\,{\rm G}\lambda$. This value is approximately $600$ times larger than the longest current EHT baselines \citep{EHT_2}. However, in the following analysis, to minimize errors from finite pixel size, we only analyze baselines shorter than $1000\,{\rm G}\lambda$.



\begin{figure*}[t]
    \centering
    \includegraphics[width=0.99\textwidth]{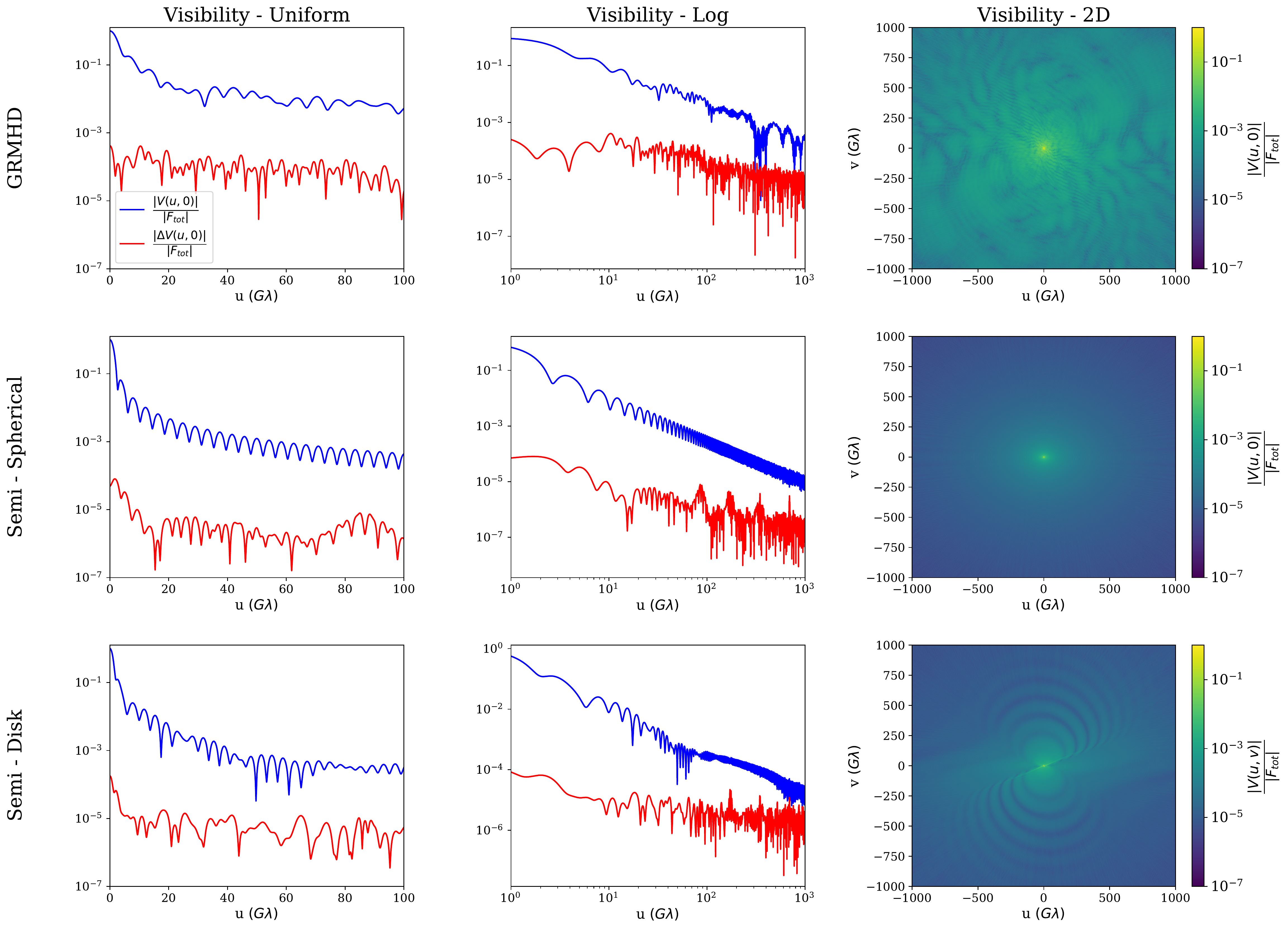}   
    \caption{Visibilities for adaptively ray-traced MAD model, spherical semianalytic model, and disk semianalytic models. Left: $|V(u,0)|$ versus $u$ on log-linear scale. Middle: $|V(u,0)|$ versus $u$ on log-log scale. Right: $|V(u,v)|$ as a 2D plot. Visibility residuals are shown in red. }
    \label{fig:madvisimages}
\end{figure*}

Figure~\ref{fig:madvisimages} shows visibility amplitudes for the adaptively ray-traced MAD model, the spherically symmetric semianalytic model, and the disk semianalytic model. Because the semianalytic images have significant emission extending beyond our specified FOV, the sharp artificial cutoffs at the edges of the FOV produce spurious high-frequency visibility power. To suppress this artificial power, we double the FOV and pixel number (keeping the image resolution fixed), and then apply a Gaussian taper with FWHM of ${\rm FOV}/4$ before computing visibilities. 
Figure~\ref{fig:madvisimages} also shows the visibility errors resulting from the adaptive ray tracing, demonstrating that these errors are a small fraction of the visibility amplitudes on all baselines. Specifically, these errors correspond to the Fourier amplitudes of the residual images defined in Section~\ref{sec:highres}. 

The MAD visibility spectrum (top row) possesses many kinks and does not display a clear-cut pattern. Although a characteristic periodicity may be evident, small-scale image power from turbulence exceeds that of the lensed emission. This turbulent power gives visibility ``noise'' that decays on long baselines approximately as $V(u) \sim u^{-p}$ with $p \approx 1.2$. 

The spherical semianalytic model (middle row), on the other hand, displays a smooth, turbulence-free spectrum.
Just as distinct subrings are not visible in the images of this model due to the spherical symmetry of the fluid distribution (e.g., \citealp{Narayan_2019}), 
distinct subrings are not visible in the visibilities of this model. 

In contrast, the disk semianalytic model (bottom row) does show clear signs of distinct photon subrings in both the images and visibilities. The spectrum falls steeply around $\sim 500\,{\rm G}\lambda$ before flattening again shortly thereafter, corresponding to the transition between the $n=1$ and $n=2$ subrings. 

\begin{figure}[t]
    \centering
    \includegraphics[width=0.45\textwidth]{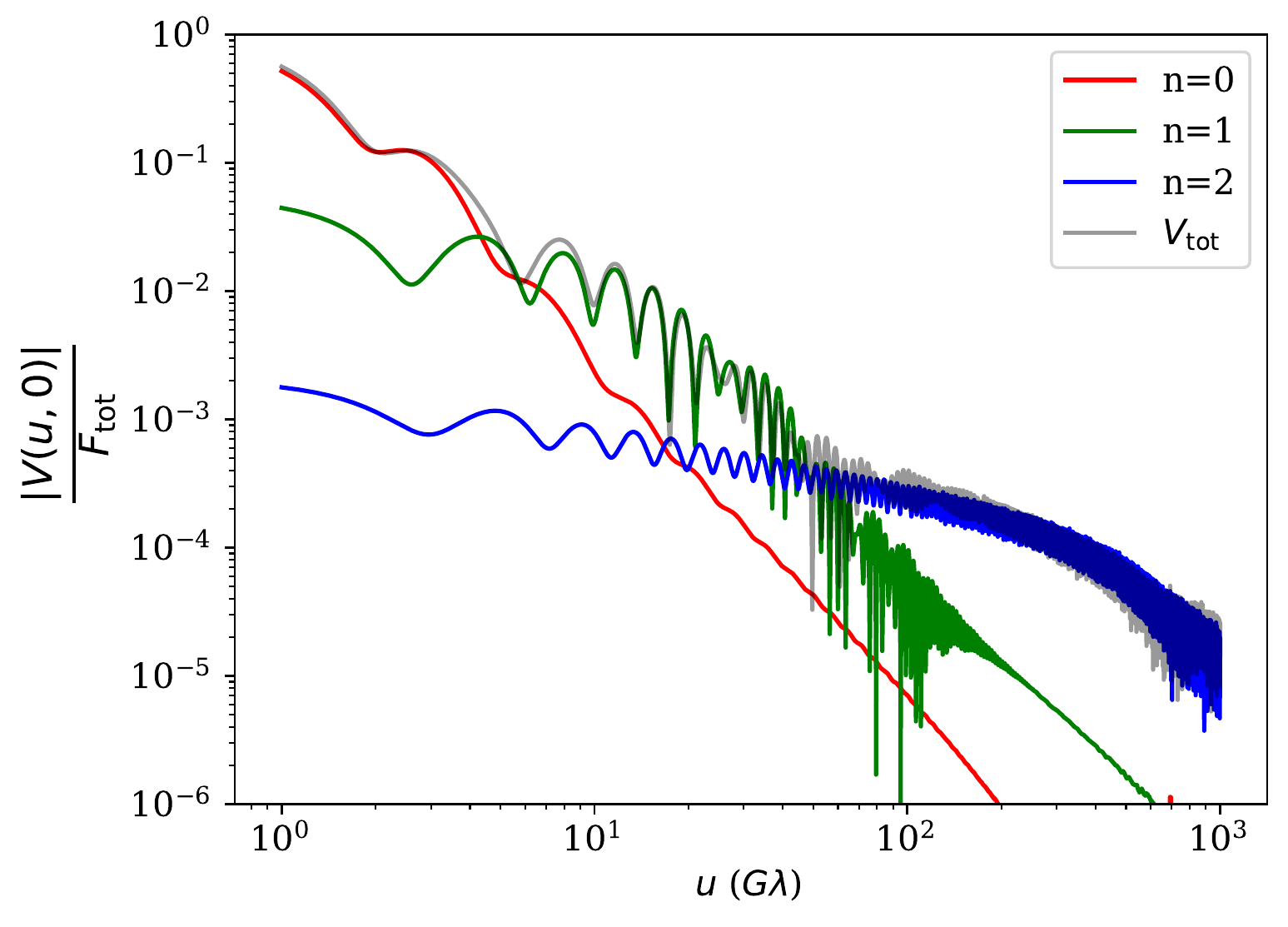} 
    \includegraphics[width=0.45\textwidth]{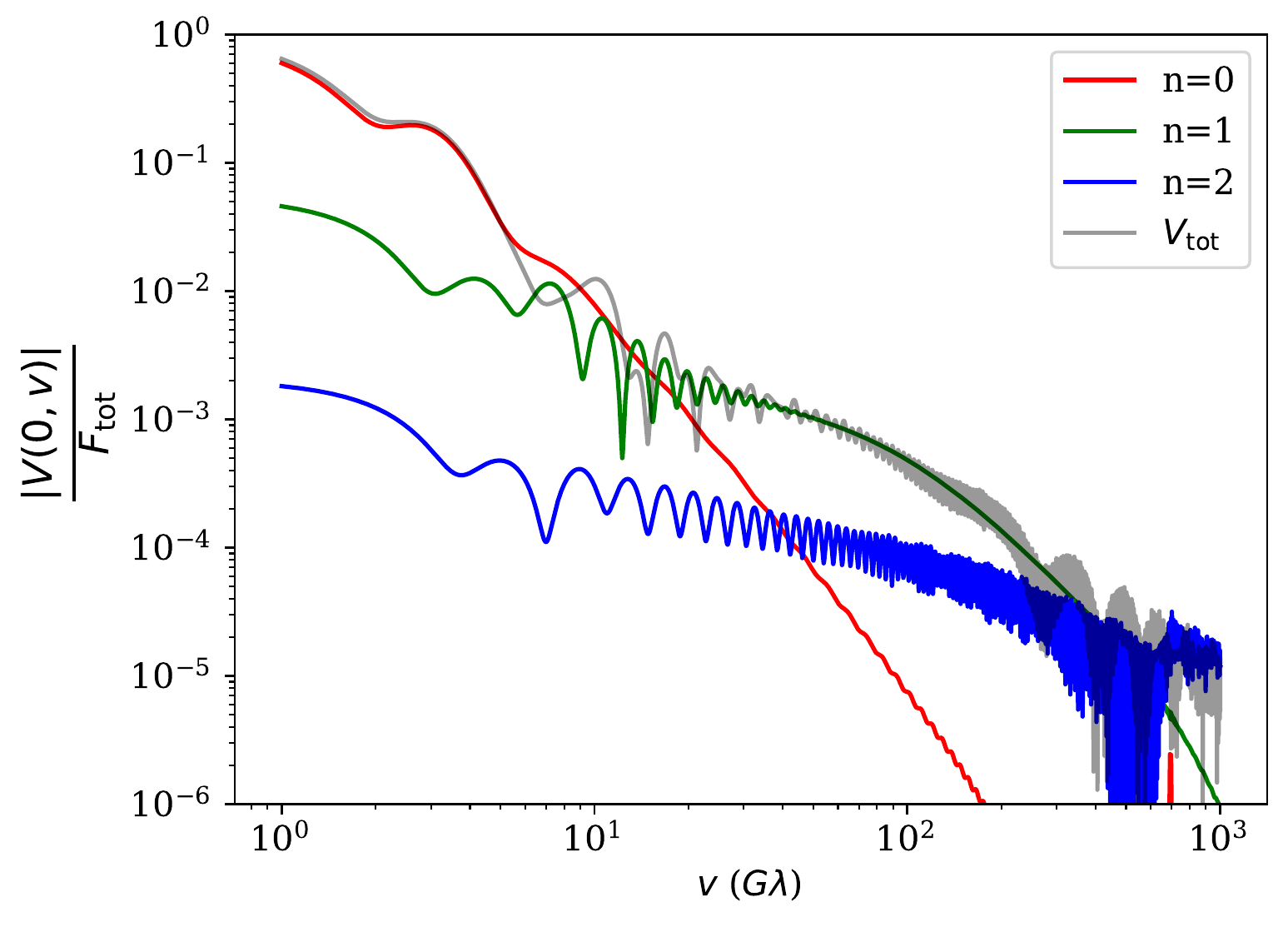} 
    \caption{Visibilities along $u$ and $v$ axes for subring-decomposed disk semianalytic model.}
    \label{fig:semisubring}
\end{figure}

\section{Applications to High Resolution Science}
\textbf{}\label{sec:apps}

In this section, we discuss specific applications of adaptive ray-tracing to larger problems in black hole simulations. In particular, we show that by generating images with extremely fine resolution, adaptive ray-tracing presents a useful tool to examine the substructure of the photon ring and to explore signatures of turbulence in simulations.

\subsection{Images of Subrings}

Resolving the photon subrings in the image domain is difficult in practice due to the exponential radial demagnification of each successive subimage \citep{Darwin_1959,Luminet_1979,Ohanian_1987}. In particular, per \citet{Johnson_2020}, the width of each subring on the screen scales as $w_n\sim w_0 e^{-\gamma n}$, where the Lyapunov exponent $\gamma$ is defined as\begin{align}
    \gamma\equiv \sqrt{\frac{2\mathcal{R}''(r_{\rm  c})}{-u_-a^2E^2}}\,K\left(\frac{u_+}{u_-}\right).
\end{align}
Here, $K$ is the complete elliptic integral of the first kind (with squared modulus $u_+/u_-$), $\mathcal{R}(r)$ is the radial effective potential for null geodesics in Kerr, $E$ is the conserved energy, $r_{\rm c}$ is the corresponding radius of spherical orbit, and $u_\pm$ denote roots of the angular effective potential. Furthermore, the radial curve $\rho_n$ of each successive subring exponentially approaches the critical curve $\rho_c$: $\delta \rho_n\approx \rho_c e^{-\gamma n}$. Hence, the image resolution required to resolve each successive subring thus grows exponentially for a fixed FOV. 

To explore properties of photon subrings explicitly, we combine our adaptive scheme with a subring decomposition code. The decomposition code can generate images corresponding to the $n^{\rm th}$ subring by only including emission from the appropriate segment of the full geodesic.
Notice that this definition of subrings differentiates between photons that may follow the same geodesic: the geodesic for a photon that makes $n$ half-orbits will overlap with the geodesic for some other photon that makes $n-1$ half orbits. Thus, a pixel that is illuminated in the $n=2$ image may also be illuminated in the $n=1$ one, and the total pixel brightness will have contributions from the second and first subrings, respectively (see, e.g., Figure 3 of \citealt{Johnson_2020}).

Figure~\ref{fig:semisubring} shows the visibility spectra of the disk semianalytic model decomposed into contributions from individual subrings. These reveal a new feature in the subring visibilities: the different subring widths between the top and bottom of the image lead to intermittent beating along the $v$ axis between the $n$ and $n+1$ subrings. When the former is resolved, the beating vanishes and the visibilities smoothly decay until reaching the level of power from the next subring.  This phenomenon does not appear on the $u$ axis, as the ring does not have significant thickness asymmetry on the horizontal axis.

\begin{figure}[t]
    \centering
    \includegraphics[width=.49\textwidth]{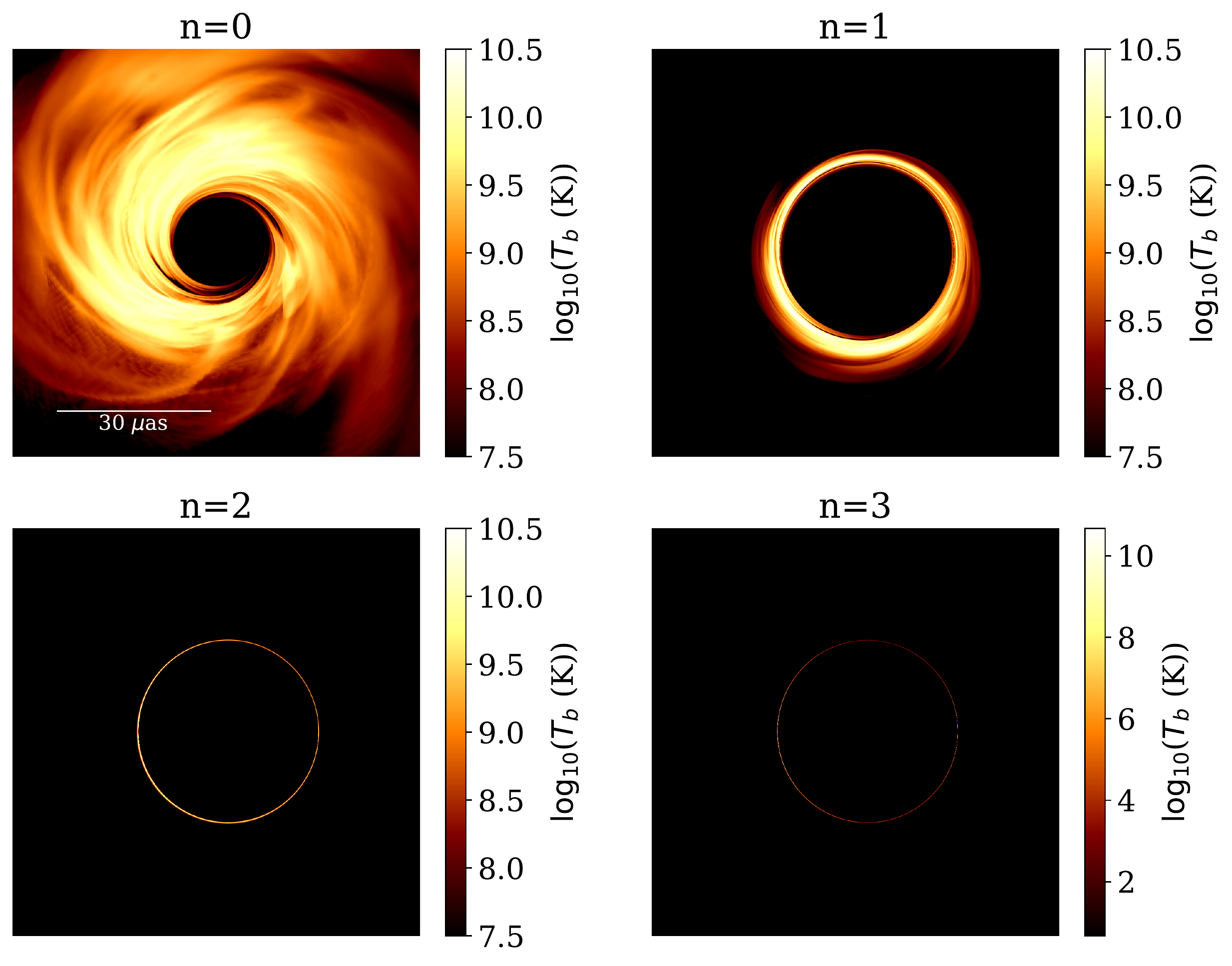}
    \includegraphics[width=.49\textwidth]{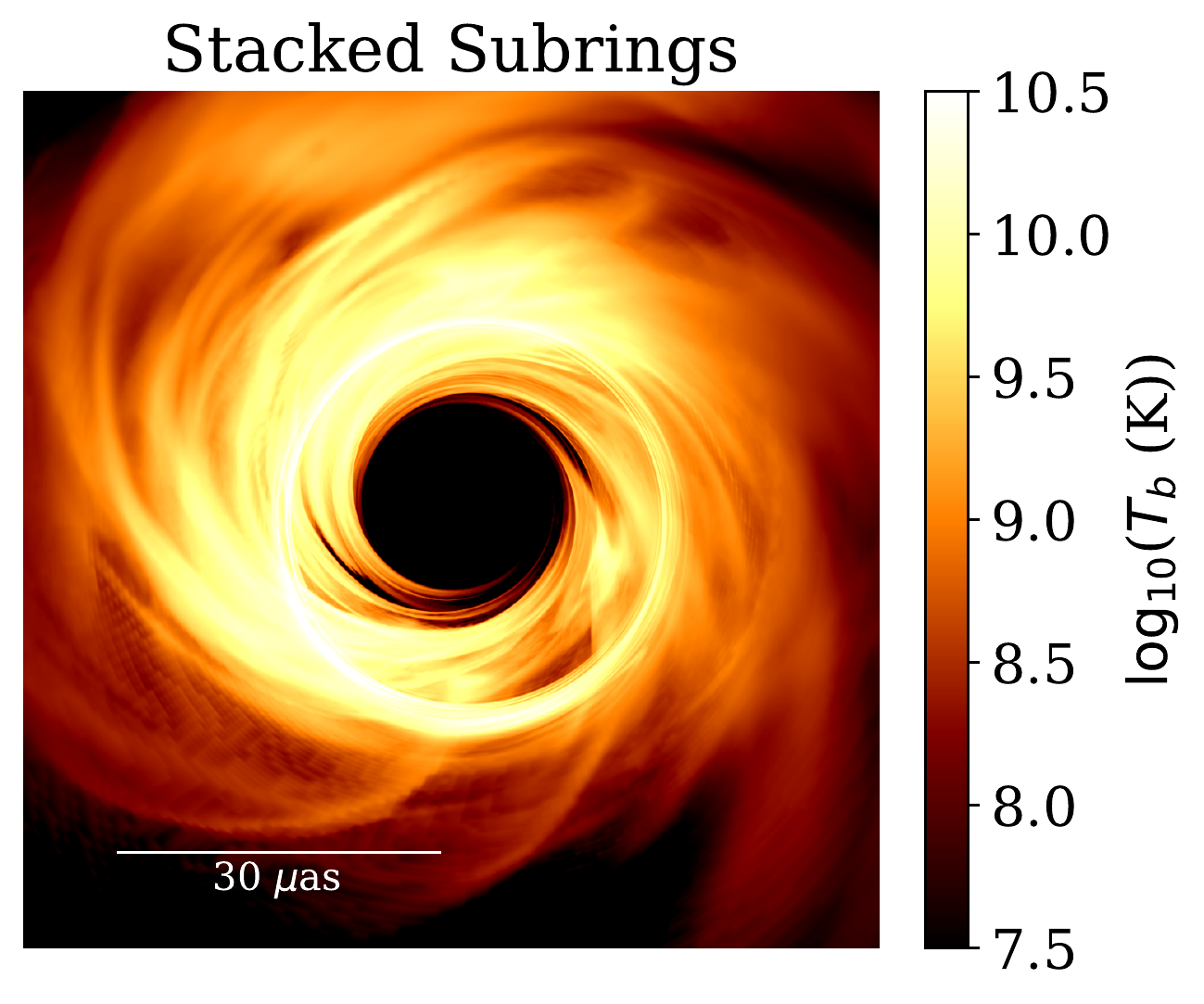}
    \caption{Adaptively ray-traced MAD model decomposed into subrings. Images for $n=1,2,$ and $3$ have resolution $32769\times 32769$ across a field of view of $80\,\mu$as, while the $n=0$ image has a resolution of $16385\times 16385$. The sum of all four subrings is shown below. All rings are visible, although the colorbar for the $n=3$ image needs a larger range of values so that we can see dim pixels. At this resolution, artifacts related to limitations in the underlying simulation are apparent in the $n=0$ image.
    }
    \label{fig:subringfig}
\end{figure}

\begin{figure*}[t]
    \centering
    \includegraphics[width=\textwidth]{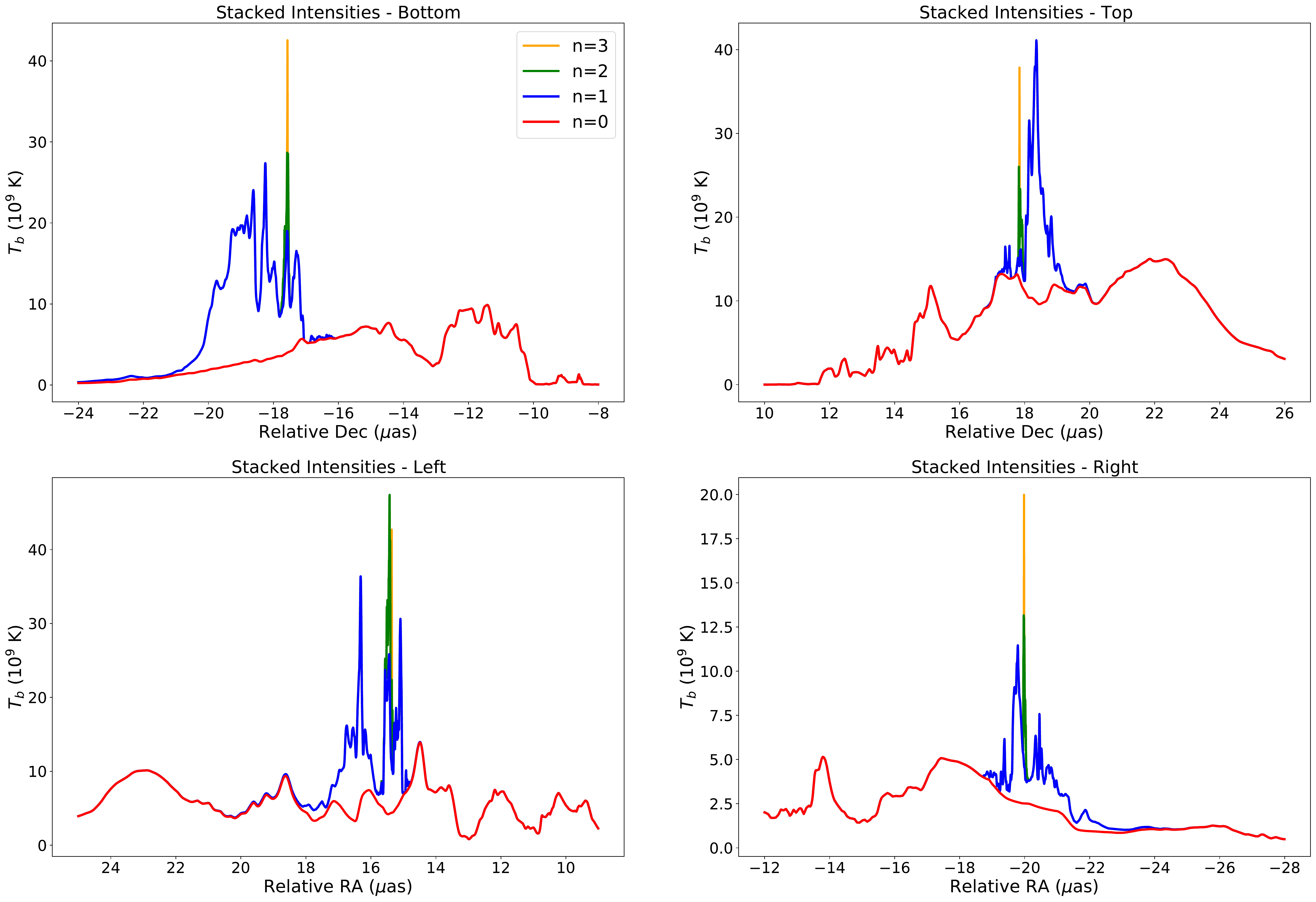}
    \caption{Stacked intensities at four locations on the ring for the same snapshot as Figure \ref{fig:subringfig}. Different colors show subrings for $n=0$ (red), 1 (blue), 2 (green), and 3 (yellow), illustrating a realization of the photon ring substructure. The contributions from individual subrings most clearly align on the bottom and right side of the image, although the right half of the image is Doppler deboosted due to the orientation of the black hole spin.}
    \label{fig:stackedrows}
\end{figure*}

To illustrate the presence of subrings at high resolutions, we again use the adpative subring decomposition code to generate $32769\times 32769$ images of the $n=1,2$ and $3$ subrings of the MAD model, now with a FOV of $80\,\mu$as. We show these images along with a lower-resolution image of $n=0$ in Figure \ref{fig:subringfig}.  Figure~\ref{fig:stackedrows} shows stacked cross-sections of the brightness profiles and zoom in on four points of the image (bottom, top, left, right). Individual subrings become thinner, with approximately constant peak brightness. Thus, the peak brightness of the sum of the first $n$ sub-images is approximately proportional to $(n+1)$.

\subsection{Time and Visibility Averaging}

We can use adaptive ray-tracing to generate high-resolution movies, which can be used to study how various averaging techniques reduce the turbulent noise present in GRMHD visibility spectra. Reducing turbulent noise is necessary to reveal universal signatures of the lensed emission surrounding the black hole, which reflect the purely geometrical properties of the spacetime. Additionally, by quantifying the amount of turbulence present in these simulations, we can obtain a better understanding of the accretion dynamics surrounding black holes and the time scales over which they vary.  

\begin{figure*}[t]
\label{fig:subringstack}
    \centering
    \includegraphics[width=\textwidth]{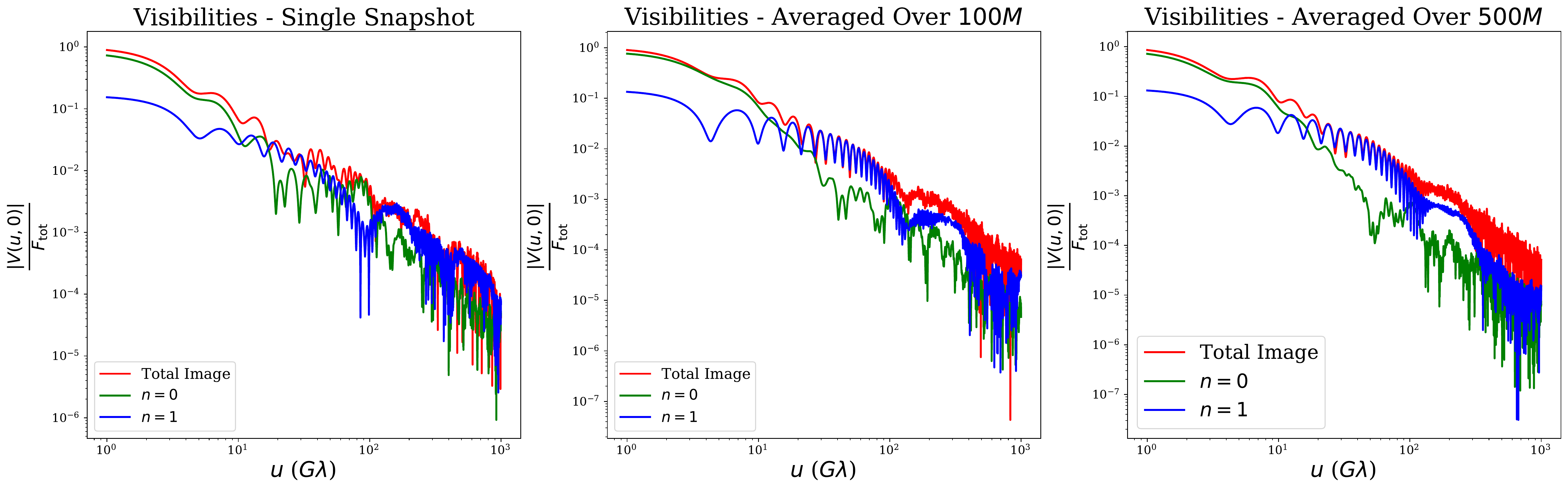}
    \caption{Single snapshot visibilities, along with time averaged visibilities for scales of $t=100M$ and $t=500M$. The image cadence for this simulation is $5M$.}
    \label{fig:vistimeavg}
\end{figure*}

In Figure~\ref{fig:vistimeavg}, we show the visibility spectra for the MAD model with $t_{\rm avg}=0 M$ (i.e., a single snapshot), as well as $t_{\rm avg}=100 M$ and $t_{\rm avg}=500M$. The images were generated with a resolution of $1025\times 1025$ and are thus capable of resolving the $n=0$ and $n=1$ subrings, whose individual visibility spectra are shown in green and blue respectively.

We see that by an averaging scale of $t_{\rm avg}=100 M$, the visibility spectrum of the full image has mostly converged to that of the $n=1$ subring for $u\lesssim 100\,{\rm G}\lambda$. However, the $n=0$ spectrum still neighbors that of $n=1$, indicating that turbulent noise is still significant. By $500M$, the $n=0$ spectrum falls below the $n=1$ spectrum in the region where we expect $n=1$ to dominate the time-averaged spectrum.

\section{Summary}
\label{sec:Summary}

In this paper, we have presented a recursive algorithm for adaptive ray-tracing, with natural applications to current and future high-resolution black hole imaging efforts. Whereas most conventional ray-tracing programs spread rays evenly across a uniform grid, our method preferentially samples rays in regions of an image with small-scale structure. 
When applied to both GRMHD and semianalytic models, we find that our algorithm reduces the time required to generate images by an order of magnitude or more. 

We then use this code to generate images with resolutions of $1025\times 1025$, $4097\times 4097$, and $32769\times 32769$ pixels. These images directly visualize the fine structure present in both the accretion flow and the photon ring, revealing the $n=1,2,$ and $3$ subrings. Finally, we explored the utility of time averaging in reducing stochastic noise in high-resolution images. 

While our algorithm reduces the computational expense required to produce high-resolution images, significant limitations in the physical modeling remain. Although we verified that the magnetization $(\sigma)$ cut negligibly alters the MSE of the adaptively ray-traced image, a sharp cut on $\sigma$ may introduce spurious high-frequency image power. Moreover, for the images generated in Section \ref{sec:apps}, we cut at $\sigma=1$, but this excludes emission near the jet in MAD models and requires additional study (see, e.g., \citealp{Chael_2019}). 

We are also fundamentally limited in resolution by the MHD cell size. For the MAD model, we use a GRMHD simulation on a spherical polar grid with a resolution of $384\times 192\times 192$ in the radial, polar, and azimuthal directions, respectively (zones are compressed exponentially toward the event horizon and lightly toward the midplane). For the SANE model, we use a resolution of $288\times 128\times 128$. The finite MHD cell size may introduce unphysical, high-power noise from sharp boundaries and will not reproduce subgrid turbulent power.

Future applications of adaptive ray-tracing could extend our results to selectively sample the time and frequency domains. For the purposes of this study, we used one frequency (230\,GHz), but fine-scale frequency structure is expected from GRMHD simulations  \citep{Ricarte_2020}. Adaptive sampling in time would allow efficient generation of high-resolution movies from numerical simulations. 

We have integrated this approach into \texttt{ipole}, but it should be compatible with any GRRT scheme that ray-traces on a rectangular grid. And while we have used unpolarized transport to generate the images in this paper, the approach generalizes to polarized images as well by simply replacing the total intensity $I(\vec{x})$ with Stokes parameters $Q(\vec{x})$, $U(\vec{x})$, and $V(\vec{x})$. Highly lensed structure near the critical curve resolved with adaptive ray-tracing may show interesting, spin-dependent symmetries in the polarization \citep{Himwich_2020}.

\acknowledgements{We thank Chi-Kwan Chan for helpful suggestions on our manuscript, and we thank Avery Broderick for useful discussions. We thank the National Science Foundation (AST-1716536, AST-1440254, AST-1935980, OISE-1743747) and the Gordon and Betty Moore Foundation (GBMF-5278) for financial support of this work. G.N.W.~was supported by a Donald C.~and F.~Shirley Jones Fellowship. This work was supported in part by the Black Hole Initiative, which is funded by grants from the John Templeton Foundation and the Gordon and Betty Moore Foundation to Harvard University. Funding for this project was provided in part by the Harvard College Program for Research in Science and Engineering.}

\appendix

\section{Estimating Relative and Absolute Errors}
Here, we derive the approximations for $\ep_{\rm abs}(\vec{x})$ and $\ep_{\rm rel}(\vec{x})$ presented in Equations (\ref{eq:epabs}) and (\ref{eq:eprel}) respectively. For convenience, we refer to the nearest neighbor errors as $\ep^{\rm NN}$ and to the linear errors as $\ep^{\rm lin}$.

\subsection{Nearest Neighbors Interpolation}
\label{sec:nearestapp}
Suppose we wish to interpolate the intensity at $\vec{x}$ directly from its nearest neighbor at $\vec{x}_1$. Because $I(\vec{x})$ and $\hati(\vec{x})$ are smooth in between $\vec{x}$ and $\vec{x}_1$, then we may apply Taylor's theorem (or in this case, just the mean value theorem) to see that\begin{align}
\label{eq:tayloreq}
    \hati-I=\nabla \hati(\vec{p})\cdot (\vec{x}-\vec{x}_1),
\end{align}
for a point $\vec{p}$ lying on the line segment connecting $\vec{x}$ to $\vec{x}_1$. To leading order in $|\vec{x}-\vec{x_1}|$, we may replace $\vec{p}$ with $\vec{x}$, giving\begin{align}
\label{eq:badnearest}
    \hati-I \approx \nabla \hati(\vec{x})\cdot (\vec{x}-\vec{x}_1).
\end{align}

Let us now restrict our attention to the rectangular gridding scheme presented in Figure \ref{fig:pixeldrawing}. On this grid, Equation \ref{eq:badnearest} is ambiguous, as $\vec{x}$ will be adjacent to multiple equidistant pixels, rendering the location of $\vec{x}_1$ ill-defined. Regardless of where we choose to set $\vec{x}_1$, however, the interpolation residual will be extremized by the quantity $|\nabla \hati(\vec{x})||\vec{x}-\vec{x}_1|$. So defining $\Delta x \equiv |\vec{x}-\vec{x}_1|$, we take\begin{align}
\label{eq:symmnearest}
    |I(\vec{x})-\hati(\vec{x})|\sim |\nabla \hati(\vec{x})|\Delta x,
\end{align}
which is the first half of Equation \ref{eq:resideq}. This expression is now symmetric -- it does not depend on which of the equidistant pixels we define to be $\vec{x}$'s nearest neighbor. 

Since we only have access to $I(\vec{x})$ at discretely sampled rays, we approximate the gradient using finite differences, which requires an examination of each of the four categories of pixel locations listed in Section \ref{sec:subdiv}. Let us suppose that $\vec{x}$ falls into Category 4, as Categories 1--3 are just simplifications thereof. The arrangement of pixels for Category 4 is shown in Figure \ref{fig:pixeldrawingzoom}. 

Since $|\nabla I|$ is rotationally invariant, we may evaluate the partial derivatives along rotated axes $x'$ and $y'$ that are aligned with the corner pixels. Then adopting the same notation as Figure \ref{fig:pixeldrawingzoom}, the central difference approximations to the derivatives are (e.g., Appendix~A of \citealp{finitedifferences})\begin{align}
    \frac{\partial \hati}{\partial x'}\bigg|_{\vec{x}}\approx \frac{\hati_4-\hati_1}{2\Delta x},  
    \qquad\qquad\qquad
    \frac{\partial \hati}{\partial y'}\bigg|_{\vec{x}}\approx \frac{\hati_3-\hati_2}{2\Delta x}.
\end{align}
Since $\hati(\vec{x})\equiv \hati_1$, the nearest neighbor error estimates become\begin{align}
    \ep_{\rm abs}^{\rm NN}(\vec{x}) \approx 
     \left|\frac{1}{\overline{I}_{\rm int}}\sqrt{\left(\frac{\hati_4-\hati_1}{2}\right)^2+\left(\frac{\hati_3-\hati_2}{2}\right)^2}\right|,
     \qquad\qquad
    \ep_{\rm rel}^{\rm NN}(\vec{x}) \approx 
     \left|\frac{1}{\hati_1}\sqrt{\left(\frac{\hati_4-\hati_1}{2}\right)^2+\left(\frac{\hati_3-\hati_2}{2}\right)^2}\right|,
\end{align}
which are the first half of Equations~\ref{eq:epabs} and $\ref{eq:eprel}$.

\subsection{Linear Interpolation}
\label{sec:linearapp}
For linear interpolation on an arbitrary 2D grid, the error estimate does not reduce to an equation as simple as \ref{eq:badnearest}. However, by restricting our attention again to the specific gridding scheme in Figure \ref{fig:pixeldrawing}, we are able to derive a straightforward error estimate as follows: 

Suppose that $\vec{x}$ falls into Category 4, as once again, Categories 1--3 will just be simplifications thereof. Adopting the same labels as Figure \ref{fig:pixeldrawingzoom}, the linear interpolation residual is explicitly given by
\begin{align}
\label{eq:explicitresid}
    \hati-I = \frac{1}{4}\left(\sum_{i=1}^4 \hati(\vec{x}_i)\right) - I(\vec{x}) = \frac{\left( \frac{1}{2}[\hati(\vec{x}_1)+\hati(\vec{x}_4)] - I(\vec{x})\right) +\left( \frac{1}{2}[\hati(\vec{x}_2)+\hati(\vec{x}_3)] - I(\vec{x}) \right)}{2}.
\end{align}
We recognize this quantity as the average of two 1D linear interpolation residuals along the $x'$ and $y'$ axes. In 1D, linear interpolation residuals scale with the second spatial derivative (see, e.g., Theorem 4.3 of \citealp{Epperson_2013}), with the leading order expansion $|\hati_{\rm 1D}-I|\approx \frac{1}{2}|\hati''(\vec{x})|\Delta x^2$. Plugging this into the numerator of \ref{eq:explicitresid}, the 2D residual becomes\begin{align}
    \hati-I=\frac{\frac{1}{2}\hati_{x'x'}(\vec{x})\Delta x^2+\frac{1}{2}\hati_{y'y'}(\vec{x})\Delta x^2}{2} = \frac{1}{4}\nabla^2 \hati(\vec{x})\Delta x^2
\end{align}
where in the last step, we used the rotational invariance of the Laplacian. This is the second half of Equation \ref{eq:resideq}.

The final task is now to approximate the Laplacian with finite differences. In doing so, we must be careful not to break the symmetry of the error approximation, and we must rely only on the pixels in Figure \ref{fig:pixeldrawingzoom}. To this end, we approximate the second derivatives by averaging the second order central differences on either side of $\vec{x}$. This gives:\begin{align}
    \frac{\partial^2 \hati}{\partial x'^2} \approx \frac{\hati_8-\hati_4-\hati_1+\hati_5}{4\Delta x^2}, \qquad\qquad\qquad 
    \frac{\partial^2 \hati}{\partial y'^2} \approx \frac{\hati_7-\hati_3-\hati_2+\hati_6}{4\Delta x^2}
\end{align}
And since $\hati(\vec{x})\equiv \frac{1}{4}(\hati_1+\hati_2+\hati_3+\hati_4)$, the error estimates become\begin{align}
    \ep_{\rm abs}^{\rm lin}(\vec{x})\approx
   \left| \frac{-(\hati_1+\hati_2+\hati_3+\hati_4)+(\hati_5+\hati_6+\hati_7+\hati_8)}{16\overline{I}_{\rm int}}\right|,
   \qquad\qquad
   \ep_{\rm rel}^{\rm lin}(\vec{x})\approx \left| \frac{-(\hati_1+\hati_2+\hati_3+\hati_4)+(\hati_5+\hati_6+\hati_7+\hati_8)}{4(\hati_1+\hati_2+\hati_3+\hati_4)}\right|,
\end{align}
which is the second half of Equations \ref{eq:epabs} and \ref{eq:eprel}.

Appropriate modifications to the scheme are made when the pixels are close to the edge of the image. In this region, we may not have access to $\hati_1-\hati_8$.

\section{Interpolation Errors for Images with a Power-Law Fluctuation Spectrum}
\label{sec:powerlaw_spectrum}

Our analysis above is appropriate for functions that are smooth and are dominated by linear or quadratic variations in a neighborhood comparable to the final pixel size. More generally, we can describe intensity fluctuations $\Delta I(\vec{x})$ by their power spectrum, $Q(\vec{u}) \equiv \left\langle \left| \Delta V(\vec{u}) \right|^2 \right \rangle$, and we can quantify expected interpolation errors statistically. By the Wiener–Khinchin theorem, the power spectrum of the intensity fluctuations is related to the two-point correlation function $C(\vec{x})$ via a Fourier transform:
\begin{align}
    C(\vec{x}) &\equiv \left \langle \Delta I(\vec{x}_0) \Delta I(\vec{x}_0 + \vec{x}) \right \rangle,\\
    Q(\vec{u}) &\equiv \left \langle \left| \int {\rm d}^2 \vec{x}\, \Delta I(\vec{x}) e^{-2\pi i \vec{u} \cdot \vec{x}} \right|^2 \right \rangle\\
\nonumber      &= \int {\rm d}^2 \vec{x}\, C(\vec{x}) e^{-2\pi i \vec{u} \cdot \vec{x}}.
\end{align}
It is also convenient to define the second-order structure function of the intensity fluctuations, 
\begin{align}
    D(\vec{x}) &\equiv \left \langle \left[ \Delta I(\vec{x}_0) - \Delta I(\vec{x}_0 + \vec{x}) \right]^2 \right \rangle\\
\nonumber      &= 2 \left[ C(\vec{0}) - C(\vec{x}) \right].    
\end{align}
For a power spectrum determined by a single, unbroken power-law, $Q(\vec{u}) \propto |\vec{u}|^{-\beta}$ and $D(\vec{x}) \propto |\vec{x}|^{\beta-2}$.

We can express interpolation errors under various schemes in terms of these functions. For instance, the root-mean-square absolute error for nearest-neighbor interpolation over a displacement $\vec{x}$ is
\begin{align}
    \epsilon_{\rm abs}^{\rm NN}\left( \vec{x} \right) &\propto \left\langle \left[ \Delta I(\vec{x}_0) - \Delta I(\vec{x}_0 + \vec{x}) \right]^2 \right \rangle^{1/2} 
  \nonumber \\&= 
    \sqrt{ D(\vec{x})}.
    \nonumber 
    \\&\propto |\vec{x}|^{\beta/2-1}
\end{align}
For comparison, if a function is smooth and is dominated by linear errors, then  $\epsilon_{\rm abs}^{\rm NN}\left( \vec{x} \right) \propto |\vec{x}|$ (see Eq. \ref{eq:resideq}). 
Hence, when $\beta < 4$, turbulent fluctuations will dominate over errors from interpolating the smooth underlying image in the limit that $|\vec{x}|\to 0$. 
On angular scales that are relevant for interpolation, black hole images are expected to show a shallower spectrum due turbulent fluctuations in the accretion flow (\citealp[see, e.g.,][]{Balbus_1991}). 
For $\beta\sim 2.5$, one has $\epsilon_{\rm abs}^{\rm NN}\left( \vec{x} \right) \propto |\vec{x}|^{1/4}$, in contrast to the factor of $|\vec{x}|$ predicted by Eq. \ref{eq:resideq}. 

We can also compare the relative error for linear and nearest interpolation strategies. For 1D linear interpolation, we have
\begin{align}
    \epsilon_{\rm abs}^{\rm lin}\left( \vec{x} \right) &\propto
    \left\langle \left[ \frac{\Delta I(\vec{x_0}-\vec{x})+\Delta I(\vec{x_0}+\vec{x})}{2}-\Delta I(\vec{x_0}) \right]^2 \right \rangle^{1/2} \nonumber
    \\&=
    \sqrt{\frac{1}{2}D(\vec{x})+\frac{1}{2}C(2\vec{x})-C(\vec{x})+\frac{1}{2}C(\vec{0})}
    \nonumber
    \\&=
    \sqrt{D(\vec{x})-\frac{1}{4}D(2\vec{x})}
    \nonumber
    \\&=\sqrt{\left(\ep_{\rm abs}^{\rm NN}\right)^2-\frac{1}{4}D(2\vec{x})}.
\end{align}       
In the case of a power-law spectrum, for the limit $|\vec{x}|\to 0$ we obtain
\begin{align}
    \frac{ \epsilon_{\rm abs}^{\rm lin}\left( \vec{x} \right) }{ \epsilon_{\rm abs}^{\rm NN}\left( \vec{x} \right) } &= \sqrt{1-2^{\beta-4}}.
\end{align}
Thus, for turbulent spectra, the improvement of linear interpolation relative to nearest interpolation is rather modest and (unlike the Taylor series analysis) is independent of the interpolated distance $\vec{x}$. For $\beta\sim 2.5$, the linear interpolation  error is only smaller than the nearest-neighbor interpolation error by a factor of $(1-2^{-3/2})^{1/2} \approx 0.8$. 

In short, small-scale turbulence in black hole accretion flows may lead to departures from the error estimates expected for a smooth image, with higher-order interpolation schemes giving less improvement than expected. For instance, while increasing the image resolution by a factor of 10 would decrease residuals by a factor of ${\sim}10$ for nearest-neighbor interpolation and ${\sim}100$ for linear interpolation of a smooth image, it may only decrease residuals by a factor of ${\sim}2$ in turbulent regions of a GRMHD image.

Unlike this statistically isotropic noise model, black hole images are restricted to a finite domain, their stochastic noise is not isotropic, and their power spectra are scale-dependent. 
The primary effect of a finite domain is to introduce correlations among different frequencies (i.e., different baselines will measure correlated fluctuations, with a correlation length given roughly by the inverse spatial extent of the image structure). The effect of position-dependent power spectra will be to blend physically distinct sources of image noise in the visibility domain. A scale-dependent power-law will give interpolation errors over an angular interval $\vec{x}$ that are primarily sensitive to the behavior near $Q(\vec{u} = \hat{x}/|\vec{x}|)$. Thus, we do not expect any of these effects to seriously modify our conclusions about interpolation errors from image stochasticity.

\software{
eht-imaging library \citep{Chael_2016},
\texttt{ipole} \citep[][]{Moscibrodzka2017ipole}, 
Numpy \citep{numpy},
Matplotlib \citep{matplotlib}
}

\bibliography{bib.bib}

\end{document}